\documentclass[aps,onecolumn,preprint,prd,superscriptaddress,nofootinbib]{revtex4-1}
\usepackage{psfig, amsmath, epsfig, amssymb,array}
\usepackage{sidecap}
\usepackage{graphicx}
\usepackage{bm}

\def\beq{\begin{equation}}
\def\eeq{\end{equation}}
\def\bea{\begin{eqnarray}}
\def\eea{\end{eqnarray}}
\def\bit{\begin{itemize}}
\def\eit{\end{itemize}}
\def\l{\left}
\def\r{\right}
\def\c{\chi}

\def\ra{\rightarrow}
\def\baa{\begin{array}}
\def\eaa{\end{array}}
\def\bt{\bar{T}}
\def\d{\partial}

\def\sq{\sqrt{2}}
\def \dblarrow#1{\stackrel{\leftrightarrow}{#1}}

\begin{document} 

\begin{titlepage}
 
\thispagestyle{empty}

\begin{flushright}UMD-PP-011-013\\V 2
\end{flushright}
\vspace{0.2cm}
\begin{center} 
\vskip .5cm
{\LARGE \bf  Improving the tunings of the MSSM by adding triplets and singlet 
}
\vskip 0.1cm 

\vskip 1.0cm
{\large Kaustubh Agashe$\, ^{a}$, Aleksandr 
Azatov$\, ^{b}$,
Andrey Katz$\, ^{a,c}$, 
and Doojin Kim$\, ^{a}$ }
\vskip 0.4cm

$^{a}${\it Maryland Center for Fundamental Physics,
     Department of Physics,
     University of Maryland,
     College Park, MD 20742, U.S.A.}
\\
$^{b}${\it Dipartimento di Fisica, Universit\`a di Roma ``La Sapienza'' and \\
INFN Sezione di Roma, I-00185 Roma, Italy}
\\
$^{c}${\it Center for the Fundamental Laws of Nature\\ 
Jefferson Physical Laboratory\\ Harvard University, Cambridge, MA 02138, U.S.A.}

\vskip 1.7cm
\end{center}

\noindent We study an extension of the MSSM which includes both new $SU(2)$ triplets with hypercharge $\pm 1$
and
a SM gauge singlet 
(a la NMSSM) 
which are coupled
to each other. We are motivated by the little hierarchy problem, 
as well as by the $\mu$ problem of the MSSM. We show that the NMSSM and the
triplet-extended MSSM 
can successfully solve problems of one another: while triplets are responsible for large correction to the lightest physical Higgs mass, the singlet's VEV explains why the $\mu$ terms 
(for the Higgs doublets and the new triplets) are naturally of order the
electroweak (EW) scale. We also show that singlet-triplet coupling significantly changes the RG evolution
of
the singlet mass squared, helping to render this mass squared negative, as required for the singlet to 
acquire a VEV. We analyze constrains on this scenario from EW precision measurements and find that a 
relatively large region of the parameter space of this model is viable, 
especially with the triplet fermions (including doubly-charged) being 
light.   
\end{titlepage}

%%%%%%%%%%
%%%%%%%%%%    Main Text
%%%%%%%%%%

\section{Introduction}

Weak scale supersymmetry (SUSY) is a
very well-motivated extension
of the standard model (SM): it naturally explains why the electroweak (EW) symmetry breaking scale is much smaller than the Planck scale, it can incorporate
a dark matter candidate, and the minimal supersymmetric SM (MSSM)
predicts the unification of the
three SM gauge coupling constants.
However, the MSSM comes with its own shortcomings: it 
cannot \emph{fully} solve the hierarchy problem since, even with the most optimistic assumptions, it is fine-tuned to the 
level of 1\% (see e.g.~\cite{Barbieri:2000gf}). This residual fine-tuning originates in a tree-level prediction that 
the SM-like Higgs in the MSSM is always lighter than the $Z$ boson. Although this result is slightly 
modified by radiative corrections, it is hard to comply with the LEP2 bounds~\cite{Barate:2003sz} on the Higgs mass, without rendering 
the stops unnaturally heavy. 

Another puzzle of the MSSM has to do with the $\mu$-term, which is required to be of 
order EW scale. Since this term is completely supersymmetric and a priori has nothing to do with SUSY-breaking, it 
is not easy to explain this coincidence of scales. While the Giudice-Masiero mechanism \cite{Giudice:1988yz} can provide 
a solution for high-scale SUSY-breaking, finding such a solution, for example, for low-scale SUSY-breaking is much harder.       

Of course, extensions of the MSSM have been proposed to 
solve these problems. For example the extension of the MSSM by addition of a (SM) gauge singlet which is 
coupled to Higgs doublets (the 
NMSSM)~(see~\cite{Ellwanger:2009dp,Chang:2008cw} for review) has been proposed to solve the
$\mu$ problem.
The idea is that a bare $\mu$-term 
is forbidden, while an effective $\mu$-term is \emph{dynamically} generated by a VEV of the singlet.
Thus the effective $\mu$-term can naturally coincide with the scale of soft SUSY breaking. However, in practice, it is typically difficult  to realize a tachyonic  singlet as is required for it to get a VEV.
It is true that the above-mentioned coupling of singlet to Higgs doublet
tends to drive the singlet mass squared negative 
in renormalization group evolution (RGE), but the up-type Higgs doublet 
mass squared is being driven negative in its own RGE from UV to IR, precisely as it happens in a regular MSSM. The tachyonic $H_u$ tends
to make the singlet mass squared more {\em positive} in
the latter's RGE mentioned above.\footnote{Of course, the resulting  tachyonic $H_u$ is otherwise
a feature (rather than a bug) since 
it results in {\em radiative} electroweak symmetry  breaking.}

Unfortunately the  NMSSM with all the couplings being perturbative\footnote{For the discussion of the NMSSM with large coupling between singlet and doublets see \cite{Barbieri:2006bg}.} up to the GUT scale also does not really ameliorate the little hierarchy problem~\cite{Mason:2009iq,Ellwanger:2006rm,Ananthanarayan:1995zr,Ananthanarayan:1996zv,Ellwanger:2005fh,Degrassi:2009yq}. This contradicts a naive expectation that the little hierarchy 
problem can be addressed in the NMSSM due to the extra Higgs quartic, which arises from the interaction with the singlet.
A reason for this ``disappointment'' is that the extra quartic coupling for the Higgs doublets, which directly contributes to the physical Higgs mass,
is suppressed in the large $\tan \beta$ limit. 
The problem is that these are 
precisely
the values of $\tan \beta$ where the tree-level
MSSM quartic coupling for Higgs doublets
(and thus the Higgs mass) tends to be
maximized. Moreover, in the NMSSM there is an additional
{\em negative} contribution to the physical
Higgs mass squared which tends to cancel the positive effect of the extra Higgs quartic. 
This effect arises 
due to singlet-doublet mass mixing term, which is
proportional to the singlet-doublet coupling and singlet VEV. The point is
that the singlet mass is also proportional to the singlet VEV so that this negative
contribution 
does not decouple
with the singlet VEV.\footnote{However, the MSSM with singlet can solve the little hierarchy 
problem if one \emph{does not insist} that the singlet gets a sufficient VEV to solve the $\mu$-problem~\cite{Delgado:2010uj}.}
Therefore if one takes the little hierarchy problem seriously, then we should consider another source for the Higgs quartic coupling, which would not decouple in the large $\tan \beta $ limit. 

One of the models which naturally possesses such a  Higgs quartic coupling is the extension of the
MSSM by the addition of $SU(2)$ triplets (dubbed TMSSM)~\cite{Espinosa:1991gr,Espinosa:1992hp,Espinosa:1998re}. This model is
especially attractive when the triplets with non-zero hypercharge
are included, such that they can couple to $H_u$ only
(unlike an NMSSM singlet or triplet with a zero hypercharge~\cite{Espinosa:1991gr,FelixBeltran:2002tb,Setzer:2006sf,
DiChiara:2008rg,DiazCruz:2007tf}).
In this case, the
extra quartic coupling for Higgs is unsuppressed
in the large $\tan \beta$ limit such that the stops significantly lighter than 1 TeV can be consistent with the LEP bounds on the Higgs mass. 
The second bug of the NMSSM in this regard
is also avoided by the TMSSM.
The  triplet VEV is required to be small, of order a few GeV (see e.g.~\cite{Flacher:2008zq})
in order to be consistent with the $\rho$ parameter, and so
is the mass mixing term
between doublets and triplets (arising in analogy to that in the NMSSM). The point is that the 
triplet
mass term 
is not proportional to its VEV,  and thus can still be large
so that the resulting negative contribution to the
physical Higgs mass squared can be 
negligible.

In this paper we propose combining these two extensions of the MSSM (namely NMSSM and TMSSM) showing that they can solve one another's problems if we couple the triplets to the singlet. Evidently the TMSSM introduces an additional $\mu$-problem (for the triplet), but this can be solved by the singlet VEV, along the lines of the solution to the usual $\mu$-problem. On the other hand, we show that the couplings of the singlet to the triplet help driving the soft mass$^2$ of the singlet negative: what is crucial here is that the triplet is not
driven tachyonic in the IR, unlike $H_u$ in the case of NMSSM mentioned above. Thus we can
solve the problem of getting suﬃciently large singlet VEV of the NMSSM.
 We also do not run into the usual problems of the NMSSM as far as the little hierarchy is concerned since the singlet interactions do not play any important role in raising the physical Higgs mass: we rely on triplets instead in achieving this goal.

Our paper is organized as follows. In section~\ref{sec:model} we present the basics of the model
in terms of parameters at the weak scale and discuss the minimization conditions of the extended Higgs potential. We
show that this model indeed addresses the little hierarchy problem such that even the tree-level mass of the Higgs can easily evade the LEP bounds. In section~\ref{sec:analysis} we perform further analysis of the model. We start by discussing the constraints on the model  from EW precision tests, mainly the $\rho$ parameter. It 
is well-known that
models with EW triplets 
in general
are subject to stringent constraints from EW precision tests since they a-priori have a new
contribution to the $\rho$ parameter from the triplet VEV. 
A neutral component of the triplets always acquires a VEV in these models. 
A trivial solution to this problem is of course to render the entire triplet (superfield) 
heavy, which has been discussed in detail in the literature. 
We concentrate instead on another part of the parameter space, where the soft masses of the triplet scalars are relatively big, but the associated $\mu$-term is small.
We show that the physical Higgs mass is larger in this region due to a suppression of
the negative contribution to it from triplet-doublet mixing driven by the triplet $\mu$-term. 
In section~\ref{sec:UV} we discuss the RG evolution of this model to higher energy scales
and discuss its implications. Finally in section~\ref{sec:conclusions} we conclude. Important RGE equations are summarized  in the appendix.

\section{The Model}
\label{sec:model}

As in any supersymmetric theory which is broken softly, the Lagrangian of the TNMSSM (which is being proposed here) 
is characterized by the superpotential, the supersymmetric gauge interactions, and the various soft breaking 
couplings (soft masses and trilinear terms).

To begin with, we consider the terms in the superpotential of the TNMSSM. As we have already  mentioned in the 
introduction, the terms in the Higgs sector depend exclusively on the gauge singlet superfield $S$, the 
$SU(2)_L$ triplet superfields $T$ and $\bar{T}$ and the MSSM Higgs doublets, $H_{ u, d }$. In addition, the model contains only dimensionless Yukawa couplings 
which will be given below, i.e., there exist no dimensionful supersymmetric parameters such as $\mu$ and $\mu_T$ ($\mu$ terms for the doublets and triplets respectively)
in the superpotential. The superpotential of the Higgs sector is given as follows: 
\begin{eqnarray}
W_{\textnormal{Higgs}}  =  S \left( \lambda H_u\cdot H_d + \lambda_T \textnormal{tr}(\bar{T} T) \right)
+ \frac{ \kappa }{3} S^3 + \chi_u H_u \cdot \bar{T} H_u + \chi_d H_d \cdot T H_d . \label{eq:model}
\end{eqnarray}
where $\lambda$, $\lambda_T$, $\kappa$, $\chi_u$, and $\chi_d$ are dimensionless Yukawa couplings. Note that 
compared to the MSSM, there is an additional physical CP violating phase coming from the superpotential: $Arg\l( \chi_u\chi_d \kappa\lambda_T^*(\lambda^*)^2\r)$. However, we defer from studying constraints from EDMs (for example) on this
new phase; instead, in this paper, we simply assume CP conservation. 

As usual, one should add the Yukawa couplings of the quark and the lepton superfields:
\begin{eqnarray}
W_{\textnormal{Yukawa}}=h_u H_u\cdot Q\bar{u}+h_d H_d\cdot Q\bar{d}+h_eH_d\cdot L \bar{e}.\label{eq:yukawasuper}
\end{eqnarray} 

Here the triplet superfields with hypercharge $Y=\pm1$ are defined as follows:
\begin{eqnarray}
T&\equiv& T^a \sigma^a =
\begin{pmatrix}
T^+/\sqrt{2}&-T^{++}\\
T^0&-T^+/\sqrt{2}
\end{pmatrix} \\
\bar{T}&\equiv& \bar{T}^a\sigma^a =
\begin{pmatrix}
\bar{T}^-/\sqrt{2}&-\bar{T}^{0}\\
\bar{T}^{--}&-\bar{T}^-/\sqrt{2}
\end{pmatrix}
\end{eqnarray}
where $\sigma^a$ are the usual $2\times2$ Pauli matrices, and the respective definitions of the products between two $SU(2)_L$ doublets and between a $SU(2)_L$ doublets and a $SU(2)_L$ triplet are given as follows:
\begin{eqnarray}
H_u \cdot H_d &=& H_u^+H_d^--H_u^0H_d^0 \\
H_u \cdot \bar{T} H_u &=& \sqrt{2}H_u^+H_u^0\bar{T}^--\left(H_u^0\right)^2\bar{T}^0-\left(H_u^+\right)^2\bar{T}^{--} \\
H_d \cdot T H_d &=&\sqrt{2}H_d^-H_d^0T^+-\left(H_d^0\right)^2T^0-\left(H_d^-\right)^2T^{++}  
\end{eqnarray}

The soft terms in the Lagrangian include:
\begin{eqnarray}
-\cal{L}_{\textnormal{soft}}
&=& m_{H_u}^2|H_u|^2+m_{H_d}^2|H_d|^2+m_{S}^2|S|^2+m_{T}^2\textnormal{tr}(|T|^2)+ m_{\bar{T}}^2\textnormal{tr}(|\bar{T}|^2) \nonumber \\
&& m_{Q}^2\left|Q \right|^2 +m_{\bar{u}}^2\left|\bar{u}\right|^2 +m_{\bar{d}}^2\left|\bar{d} \right|^2 +m_{L}^2\left|L \right|^2 +m_{\bar{e}}^2\left|\bar{e} \right|^2 \nonumber \\
&&(A_{h_u}Q\cdot H_u\bar{u}-A_{h_d}Q\cdot H_d\bar{d}-A_{h_e}L\cdot H_d \bar{e}  \nonumber \\
%%%%%%%%%%%%%%%%%%%%%%%%%%
&&  A S H_u\cdot H_d + A_T S \ \textnormal{tr}(T\bar{T}) + \frac{A_{\kappa}}{3}S^3+ A_u H_u\cdot\bar{T}H_u+A_d H_d\cdot T H_d+h.c.) \label{eq:new}
\end{eqnarray}
Note that, in addition to R-parity, we explicitly imposed a $\mathbb{Z}_3$ symmetry which forbids bare $\mu$ and $B\mu$ terms both for Higgs doublets and triplets\footnote{For the study of NMSSM without $\mathbb{Z}_3$ symmetry see \cite{Ross:2011xv}.}. The effective $\mu, \ B\mu$ terms form when $S$ gets a VEV. We also assume that   
all $A$-terms are small for simplicity, and thus one can expect a light pseudoscalar, the $R$-axion, which will be discussed in detail in section~\ref{sec:raxion}. This assumption is also well motivated in context of flavor-safe mediation 
mechanisms for SUSY breaking, for example, gauge mediation.

As advertised in the introduction, one can briefly see that this model, i.e., combining the singlet and triplet
extensions of the MSSM, can solve both ``$\mu$-problem(s)'' and ``little hierarchy problem''. For the $\mu$-problem(s), like in the NMSSM, a vacuum expectation value $v_s$ of singlet of the order of the weak or SUSY breaking scale will generate an effective $\mu$-term for the Higgs doublet and the triplet with 
\beq \label{eq:effmu}
\mu^{\textnormal{eff}} = \lambda v_s  \ \ \ 
\mu^{\textnormal{eff}}_T = \lambda_T v_s. 
\eeq
For the little hierarchy problem, clearly one can see that the coupling of triplet to \textit{up}-type Higgs $H_u$ introduces the extra quartic couplings for Higgs without any mixture with \textit{down}-type Higgs $H_d$ in the Higgs (tree-level) potential:
\begin{eqnarray}
V_{\textnormal{Higgs}} \ni\;\sim \chi_u^2(H_u)^4.
\end{eqnarray}
In the next section, we will see that this leads to an enhancement of SM-like Higgs mass even in the large tan$\beta$ limit. 

%%%%%%%%%%%%%%%%%%%%%%%%%%%%%%%%%%%%%%%%%%%%%%%%%%%%%%%%%%%%%%%%%%%%%%%%%%%%%%%%%%%%%%%%%%%%%%%%%%%%%%%%%%%%%%%%%%%%%%%%%%%%%%%%
\subsection{SM-like vacuum of TNMSSM}
\label{sec:vacuumstructure}
Plugging the VEVs into the full Higgs potential one gets:
\begin{eqnarray}
V_{\textnormal{Higgs}}&=&(2\chi_u v_u v_{\bar{T}}+\lambda v_s v_d)^2+(2\chi_d v_d v_T + \lambda v_s v_u)^2+(\kappa v_s^2-\lambda v_u v_d - \lambda_T v_T v_{\bar{T}})^2 \nonumber \\
&& +(\chi_u v_u^2+\lambda_T v_s v_T)^2+(\chi_d v_d^2+\lambda_T v_s v_{\bar{T}})^2+\frac{g^2+{g'}^2}{8}(v_u^2-v_d^2+2v_T^2-2v_{\bar{T}}^2)^2 \nonumber \\
&& + m_{H_u}^2v_u^2+m_{H_d}^2v_d^2+m_S^2v_s^2+m_T^2v_T^2+m_{\bar{T}}^2v_{\bar{T}}^2 \nonumber \\
&&-2Av_s v_u v_d-2A_T v_s v_T v_{\bar{T}} +\frac{2}{3}A_{\kappa}v_s^3-2A_u v_u^2 v_{\bar{T}} -2A_d v_d^2 v_T \label{eq:min}.
\end{eqnarray}
where $g'(\approx 0.35)$ and $g$ denote $U(1)_Y$ and $SU(2)_L$ gauge couplings, 
respectively.  
In the TNMSSM, the mass of the $Z$ boson has the same form as in the MSSM, but the electroweak symmetry breaking (EWSB) VEV 
for the doublets are modified due to the presence of triplet VEVs:
\begin{eqnarray}
M_Z^2&=&\frac{{g'}^2+g^2}{2}v^2 \equiv \hat{g}^2v^2  \\
v^2&=& v_u^2+v_d^2+4v_T^2+4v_{\bar{T}}^2 \approx (174 \textnormal{GeV})^2,
\end{eqnarray} 
and tan$\beta$ is defined by the ratio of $v_u$ to $v_d$ as usual: $\tan \beta \equiv {v_u}/{v_d}$.\footnote{Given that in our case $\left( v_u^2 + v_d^2 \right)$ does not sum to the measured value of the SM Higgs VEV$^2$
(unlike in the MSSM or even the NMSSM),  this definition might be somewhat misleading. Nonetheless, since we know that the corrections from the triplet VEVs are small (as required by the $T$-parameter), we will still loosely use this definition.}

Since we introduced  a singlet and two triplets, we have five minimization equations including the ones for usual up- and down-Higgses. 
In general, the vacuum expectation values for the triplets must be small to avoid large $\rho$ parameter correction, which will be fully investigated in section~\ref{sec:eptriplet}. Assuming small VEVs for the triplets, i.e., $v_T,\;v_{\bar{T}}\approx 0$, one can easily derive the following relation for the ratio of the $v_u, v_d$ using minimization equations for $H_u$ and $H_d$:  
\begin{eqnarray}
\frac{v_uv_d}{v_u^2+v_d^2}=\frac{1}{2}\sin 2\beta =\frac{v_s(\lambda \kappa v_s+A)}{2(\chi_u^2v_u^2+\chi_d^2v_d^2)+\lambda^2(2v_s^2+v_u^2+v_d^2)+m_{H_u}^2+m_{H_d}^2}
\end{eqnarray}
Note that this relation reduces to the usual NMSSM relation in the limit $\c_u,\c_d=0$ as expected. In order to have non-zero $v_u$ and $v_d$ the numerator should not vanish, i.e., $\lambda \kappa v_s +A \neq 0$. Indeed, the TNMSSM accommodates such a non-zero numerator. To see this, one notices that a large $v_s$ is required in order to generate a sufficiently large effective $\mu$-term (see Eq.~(\ref{eq:effmu})) like the case of the NMSSM~\cite{Ellwanger:2009dp}. In addition, since small $A$-terms are assumed as mentioned before, this condition for non-vanishing $v_u$ and $v_d$ is readily attained in most of the parameter space.

One can easily find that the minimization condition for $v_s$ reduces to the corresponding NMSSM-like form under the assumption of small VEVs for the triplets~\cite{Ellwanger:2009dp}:
\begin{eqnarray}
(2\kappa^2 v_s^2+\lambda^2(v_u^2+v_d^2)-\kappa \lambda v_uv_d+A_{\kappa}v_s+m_S^2)v_s=0
\end{eqnarray}
One can see that for small $A_\kappa$ and $v_s\gtrsim v_{d}$, one should demand $m_S^2<0$. Here we simply assume
this condition, but we show in section~\ref{sec:UV} that our model can naturally have this feature. 

%%%%%%%%%%%%%%%%%%%%%%%%%%%%%%%%%%%%%%%%%%%%%%%%%%%%%%%%%%%%%%%%%%%%%%%%%%%%%%%%%%%%%%%%%%%%%%%%%%%%%%%%%%%%%%%%%%%%%%%%%%%%%%%%%%%%%%5
\subsection{Higgs mass in the TNMSSM}
\label{sec:higgssector}
To begin the discussion of the SM-like Higgs mass in the TNMSSM, one may 
consider a few interesting limits, depending on the hierarchy between the soft mass term and the $\mu$-term for triplets. As we will discuss in more detail in section~\ref{sec:eptriplet}, one is required to have large masses for the triplets to avoid a significant  correction
to the $\rho$-parameter. To obtain large triplet scalar masses, either a soft mass for triplets, or a $\mu$-term for triplets, or both of them should be large enough (again see section~\ref{sec:eptriplet} for details) so that we will discuss three distinct cases. One of them has large $\mu_T$  and small triplet soft masses, which implies that the triplets can be integrated out in the supersymmetric limit. The opposite regime is the highly non-supersymmetric one, where $m_T \gg \mu_T$, and we will find it to be the most interesting phenomenologically. One can think about the third regime, $m_T \sim \mu_T$ as a kind of an intermediate case.   

It is easy to estimate the Higgs mass in the limiting cases, and the intermediate case can be understood as the admixture of the two extreme limits. Let us begin with the limit where the soft mass is dominant and the $\mu_T$ is small.   
The tree level Higgs mass spectra, in general, can be obtained by diagonalizing the associated mass mixing matrices. To find the mass of the SM-like Higgs we have to (typically) find the lightest eigenvalue of the ($5\times 5$) mass matrix of the CP-even neutral scalars. The associated calculations will be made numerically in our parameter scans, but in order to develop some intuition we can look at the following analytical upper bound on the mass of the lightest Higgs~\cite{Espinosa:1991gr}:
 \begin{eqnarray}
m_{h^0}^2\leq M_Z^2 \left(\cos^2 2\beta+\frac{\lambda^2}{\hat{g}^2}\sin^22\beta+\frac{\chi_d^2}{\hat{g}^2}\cos^4\beta+ \frac{\chi_u^2}{\hat{g}^2}\sin^4\beta\right) \label{eq:higgsupperbound}
\end{eqnarray}
Note that the last term in Eq.~(\ref{eq:higgsupperbound}) is proportional to $\sin^4 \beta$ (originating from the coupling of only the up-type Higgs to triplet) which is good for solving the little hierarchy problem since it is maximized
precisely where the SM-like Higgs mass is in the MSSM, i.e., $\beta \rightarrow \frac{\pi}{2}$.  
On the contrary, as mentioned in the introduction, in the NMSSM, the enhancement of the Higgs mass at the tree level is suppressed in 
this large $\tan \beta$ limit [see 2nd term
in Eq. (\ref{eq:higgsupperbound})] .
Also note that the bound Eq.~(\ref{eq:higgsupperbound}) is saturated in the limit of small mixing between doublets and singlet
or triplets.  Such mixing arises from various $F$-terms and reduces the mass of the Higgs below the 
bound in Eq. (\ref{eq:higgsupperbound}).
In the case of NMSSM \cite{Ellwanger:2009dp}, the doublet-singlet mixing
from the $F$-term of $H_d$ then results in the
following estimate for the Higgs mass (in the large $\tan\beta$ limit):
\bea
m_{h^0}^2\sim M_Z^2-\frac{\lambda^4v^2}{\kappa^2}
\eea

Now let us look at the effects coming from the doublet-triplet mixing in the TNMSSM. The effects of doublet-triplet mixing from $F$-terms of singlet and $H_u$ are
proportional to the VEV of the triplet. Since bounds on the $T$ parameter require
the VEV of the triplet to be very
small (see discussion in section \ref{sec:eptriplet}), this mixing is negligible.
Similarly, the doublet-triplet mixing from the $F$-terms of $H_{ u, d }$ (which is proportional to $\mu$-term for doublets) is
suppressed in the large $\tan \beta$ limit (which is our interest here). 
Thus, the most important contribution to the doublet triplet mixing arises from the $F$ term of the triplet
\begin{eqnarray}
| F_T | ^2 & \sim & 
\chi_u \mu_T \bar{T}^{ \dagger } H_u H_u + h.c,
\end{eqnarray}
which results in a shift in the Higgs mass:
\begin{eqnarray}
\delta m_{ h^0 }^2 & \sim & - \frac{ \left( \chi_u \mu_T v_u \right)^2 }{ \mu^2_T + m_T^2 }
\label{mu-driven}
\end{eqnarray}
However in the limit $m_T \gg \mu_T$  this effect (which does not
depend on the triplet VEV) is suppressed and we can estimate the overall correction to the Higgs mass to be
\bea
m_{h^0}^2\sim M_Z^2-\frac{\lambda^4v^2}{\kappa^2}+\chi_u^2 v^2
\eea
We see that by choosing appropriate values of the couplings $\lambda$, $\chi_u$, and $\kappa$ (for example,
$\chi_u \sim 0.5 > \lambda \sim \kappa$),
we can easily be above the LEP2 bound on the Higgs mass (even at tree-level).

On the other hand, the estimate of the Higgs mass in the opposite limit, where $\mu_T$ is dominant, is different. 
In this case, the shift in Higgs mass given in Eq.~(\ref{mu-driven})
actually cancels the last term in Eq.~(\ref{eq:higgsupperbound}).\footnote{See reference \cite{DiChiara:2008rg} for a similar
discussion in the model with zero hypercharge triplet.}
An equivalent way to see this cancellation is that, 
in this limit, the triplets must be supersymmetrically integrated out, and thus their remnant effects appear via non-renormalizable effective superpotential~\cite{Brignole:2003cm,Dine:2007xi,Carena:2009gx, Carena:2010cs}. After the 
suspersymmetric integrating out is performed, one finds no triplet 
contribution in the Higgs mass proportional to $\sin^4 \beta$, but rather 
\bea
\delta m_{h^0}^2 \; (\hbox{triplet only}) \simeq \frac{\mu}{\mu_T}\chi_u\chi_d v^2 \sin 2\beta=\frac{\lambda}{\lambda_T}\chi_u\chi_d v^2 \sin 2\beta + \ldots 
\label{SUSYtriplet}
\eea 
where the dots stand for corrections, coming from  hard SUSY breaking terms, proportional to powers of $m_T / \mu_T$.
Clearly, this contribution to the SM-like Higgs mass favors moderate $\tan\beta$ (i.e., $\beta \sim \pi/4$), where the MSSM contribution is not saturated while the NMSSM contribution is saturated. 

It is interesting to compare the size of the correction to the Higgs mass due to the triplets only (i.e.,
ignoring the singlet) in SUSY versus non-SUSY limit.
\bea
R\equiv \frac{(\delta m_{h^0}^2)_{\hbox{SUSY}} }{(\delta m_{h^0}^2)_{\hbox{non-SUSY}}} \sim \frac{\frac{\lambda}{\lambda_T}\chi_u\chi_d v^2 \sin 2\beta}{M_Z^2\left(\frac{\chi_d^{ \prime \; 2 } }{\hat{g}^2}\cos^4\beta'+\frac{\chi_u^{ \prime \; 2 }}{\hat{g}^2}\sin^4\beta'\right)}=\frac{r_{\lambda}r_{\chi}\sin 2\beta}{r_{\chi}^2\cos^4\beta'+\sin^4\beta'}
\eea   
where $r_{\lambda}=\lambda/\lambda_T$ and $r_{\chi}=\chi_d \chi_u / \chi_u^{ \prime \; 2 } $. To compare the maximum contributions in the respective limits, one should take $\beta \rightarrow \pi/4$ and $\beta' \rightarrow \pi/2$ so that $R$ is simplified to $R=r_{\lambda}r_{\chi}$. 
Since in the SUSY limit, $\mu_T$ is larger than the weak scale (whereas
$\mu$ is at the weak scale for naturalness), we have $r_{ \lambda } \ll 1$. So, unless we choose the $\chi$-type couplings
in the two limits so that
$r_{ \chi } \gg 1$, we see that 
it is difficult to get 
larger size of Higgs mass correction by introducing the triplets in the SUSY limit than that in the non-supersymmetric limit. 

\begin{figure}[t]
\centering
\includegraphics[width=7.5cm]{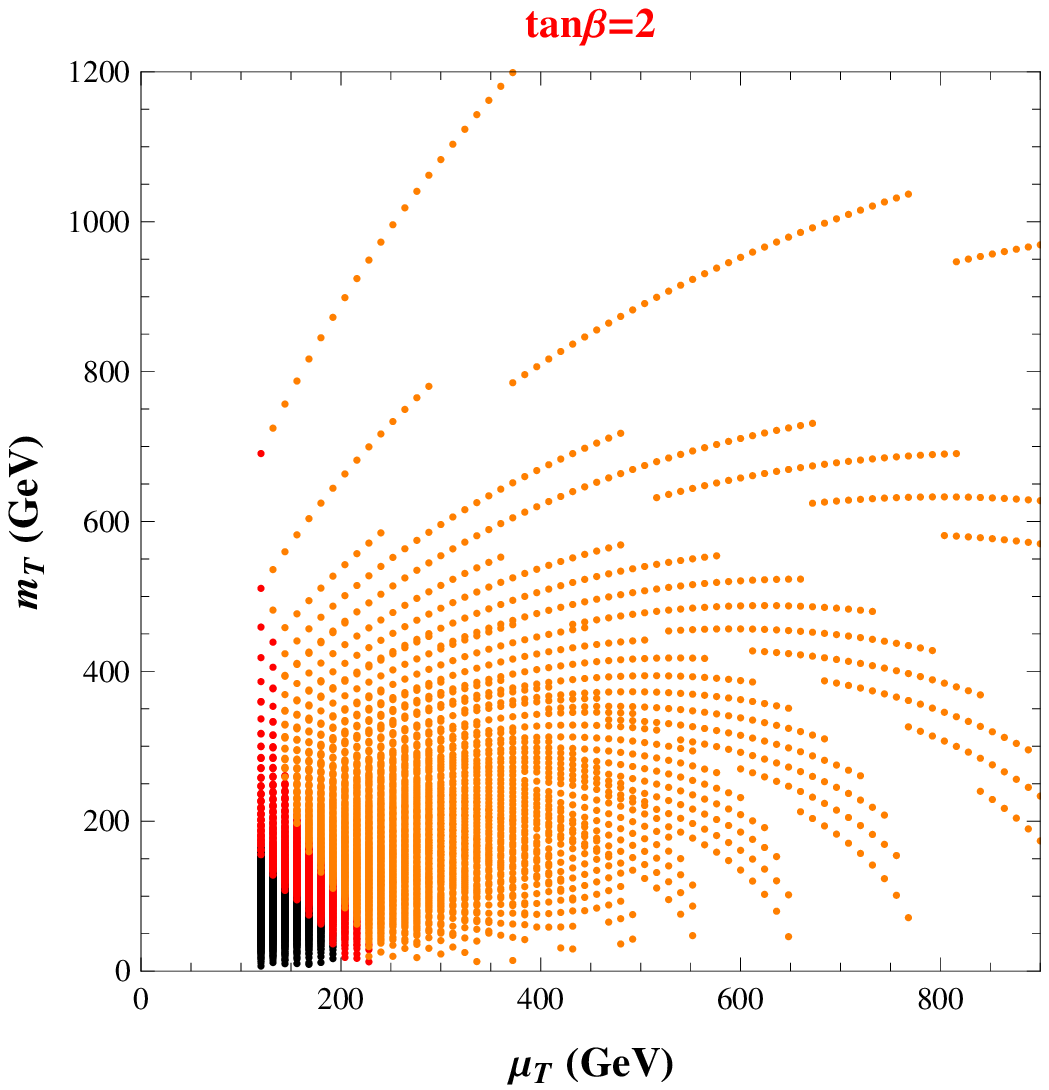}
\hspace{0.2cm}
\includegraphics[width=7.5cm]{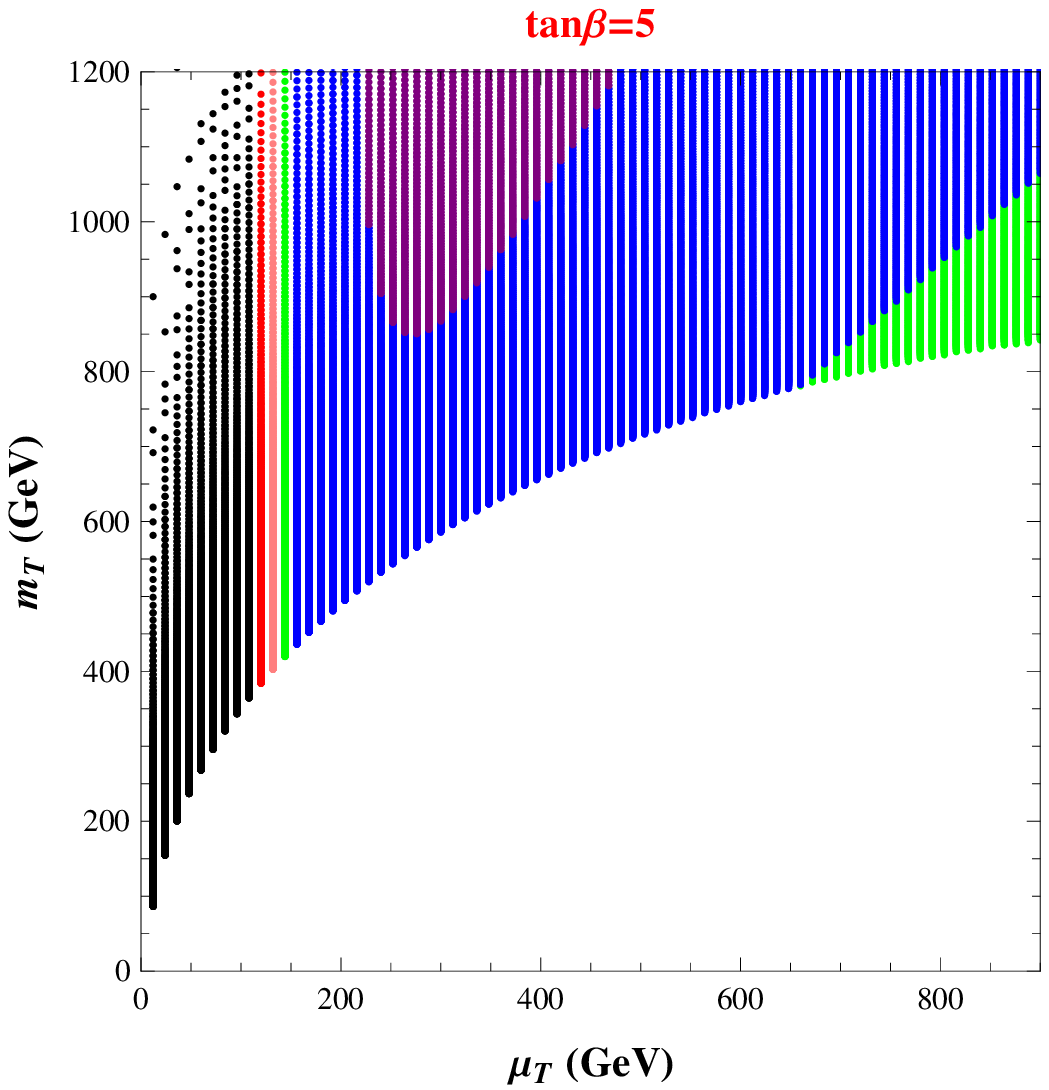}
\caption{Contour plots of the SM-like Higgs mass (including one-loop radiative correction from stop)
in the plane of $m_T\approx m_{\bt}$ and $\mu_T$. The relevant parameters (see text
for an explanation) are chosen
as follows:  
$\lambda=0.30,\;\kappa=0.30,\;\lambda_T=-0.60,
\;\chi_u=-0.30,\chi_d=-0.55$, $m_{\tilde{t}}=300$ GeV
and all $A$-terms are assumed to vanish.
The black, the red, the orange points correspond to the Higgs mass in the ranges $[0,110]$ GeV, $[110,114.4]$ GeV, and $[114.4,120]$ GeV, respectively. 
For the remaining colors (in the order pink, green, blue, and purple), the upper endpoints of this range are
increased successively by 5 GeV, i.e., pink is $[120, 125]$ GeV and so on. Note that the large blank region on the plot with $\tan\beta=5$ is due to the  constraint from the $\rho$ parameter. This constraint is weakened for the case of $\tan\beta=2$ because of an accidental cancellation between $F$-terms 
of the $\bt$ and $H_d$ which results in a smaller VEV for the triplet.
The 
vacant regions
appearing 
for 
large soft masses are an artifact of the numerical calculation.
}
\label{fig:contour}
\end{figure}
 
Figure~\ref{fig:contour} is a contour plot of SM-like Higgs mass in the TNMSSM in the plane of triplet soft mass term versus $\mu$-term for the triplet.  For the scan we take one-loop correction from stop into consideration \cite{Martin:1997ns},
\bea
\delta m_h^2\sim \frac{3}{4\pi^2}\sin^2\beta y_t^2m_t^2 \ln \l(\frac{ m_{\tilde t_1} m_{\tilde t_2}}{m_t^2}\r)
\eea
 neglecting any stop mixing in order to be conservative. Namely, the mixing between $\tilde t_R$ and $\tilde t_L$ pushes the Higgs mass even higher so that we are generically underestimating the Higgs mass. 
We
show only the points which respect the $\rho$-parameter constraint which will be discussed in detail in section~\ref{sec:ewtmssm}. Note that we completely neglect the $A$-terms for all of the Higgs fields in this scan. From this
figure, we see that (as expected from the estimates above) it is easier to get large enhancement of the physical Higgs mass (namely $m_h > 120 $ GeV) when $\tan\beta$ is large {\em and} when there is a large hierarchy between the soft triplet mass term and the $\mu$-term for triplet. It does not necessarily mean that moderate $\tan \beta $ is excluded by the Higgs mass limit, but getting
$m_h > 120 $ GeV in this case is difficult.

\subsection{Remarks about mass spectrum}
Analysis of a full mass spectrum (i.e., including squarks and sleptons) is highly model dependent and heavily relies on underlying assumptions about the 
mediation scheme. Here we just make several comments regarding the spectrum of EW Higgses, charginos and neutralinos, which 
directly follows from our previous considerations of the Higgs sector. 

We begin with the spectrum of the neutral scalars and pseudo-scalars.
One can easily derive the relevant mass matrix elements 
$(M_0^{2})_{ij}$ in the basis of ($ H_u^0, H_d^0,T^0,\bar{T}^0,S$). 
We notice that in the limit of the vanishing  $A$ terms and gaugino masses our Lagrangian is invariant 
under extra $U(1)_R$ symmetry under which all the chiral fields carry a charge 2/3. 
This leads to the appearance of the light pseudo-scalar, i.e., the
pseudo-Nambu-Goldstone boson associated with the spontaneous breaking 
of this symmetry. More detailed discussion of this ``$R$-axion'' is left to section~\ref{sec:raxion}.

Moving onto charged scalar particles, in the TNMSSM these are of two types: singly charged and 
doubly charged. In the singly charged Higgs sector, i.e., $(M_{\pm}^{2})_{ij}$, the basis
consists of
$H_u^+, H_d^{-*},T^+,\bar{T}^{-*}$. 
For the doubly charged Higgs sector, there are contributions only from triplets (i.e., $T^{++}$ and 
$\bar{T}^{--*}=\bar{T}^{++}$), and thus the associated mass matrix $\l(M_{2\pm}^{2}\r)_{ij}$ is simply a $2\times 2$ matrix. 

Similarly, the fermionic mass matrices can be constructed. The EW gauginos contribute in the neutral and singly-charged sectors, but not in the doubly-charged sector 
(where only the triplet contributes, just like for scalars above).
 
\begin{figure}[ht]
\centering
\includegraphics[width=4.7cm]{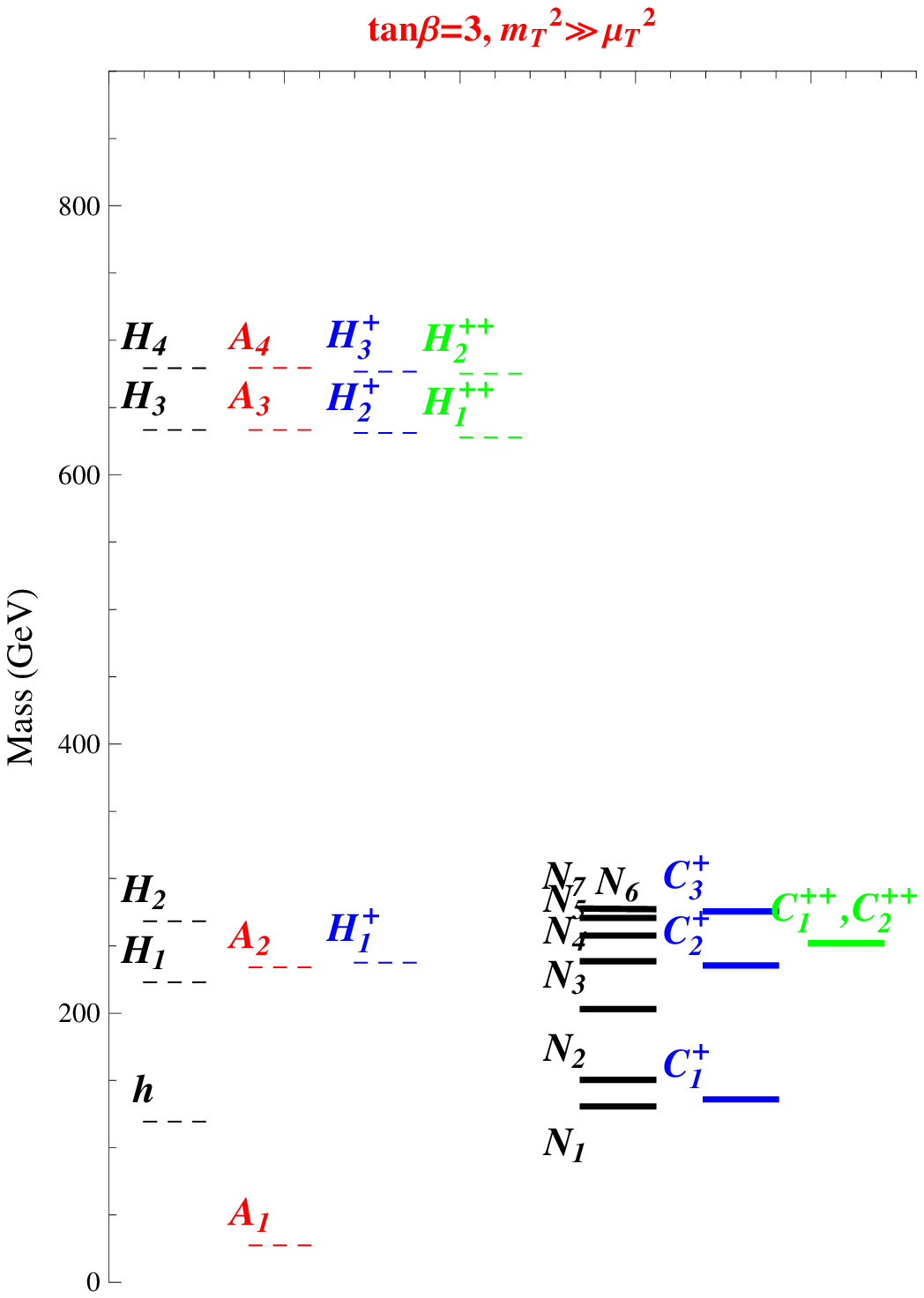}\hspace{0.2cm}
\includegraphics[width=4.7cm]{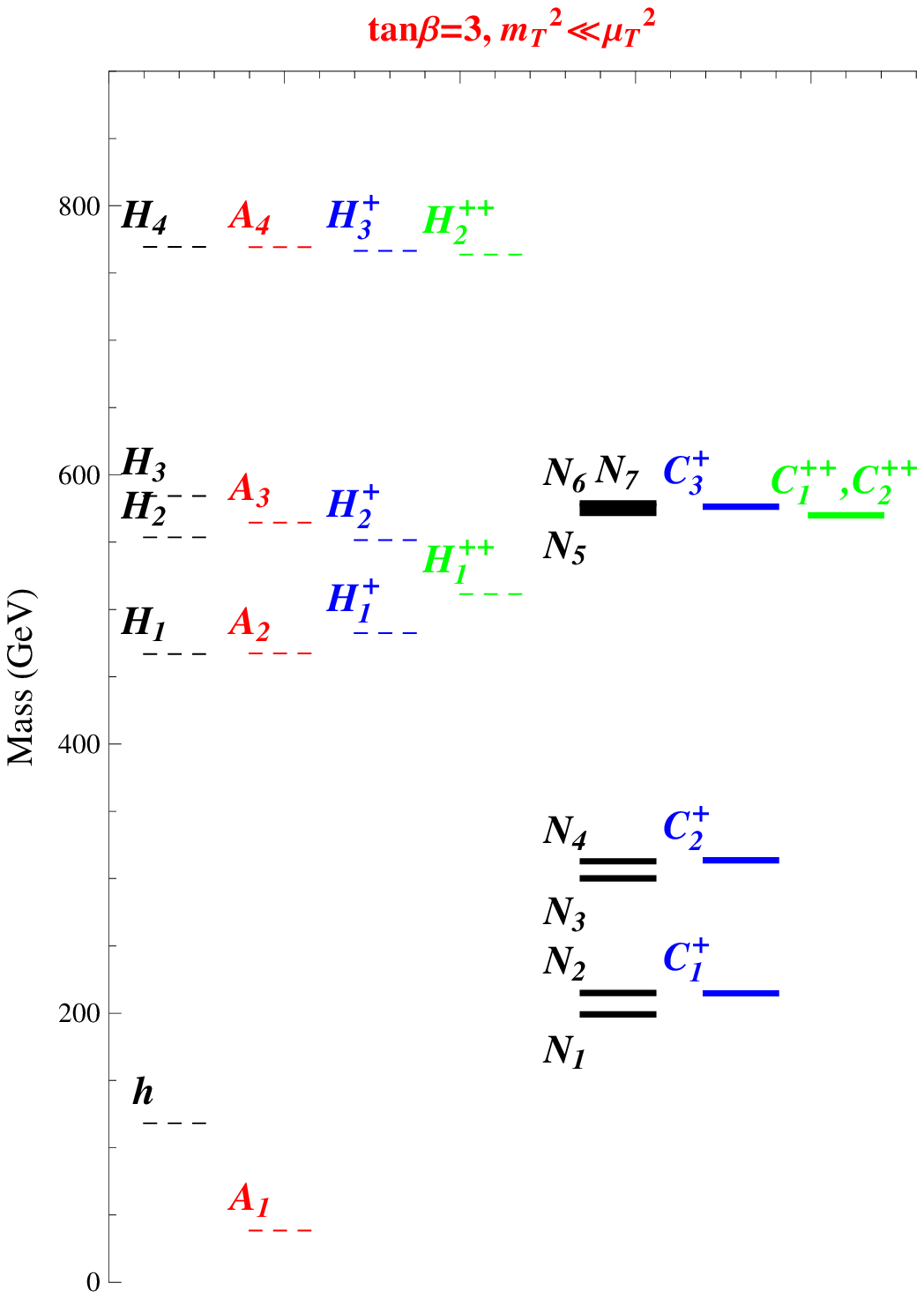}\hspace{0.2cm}
\includegraphics[width=4.7cm]{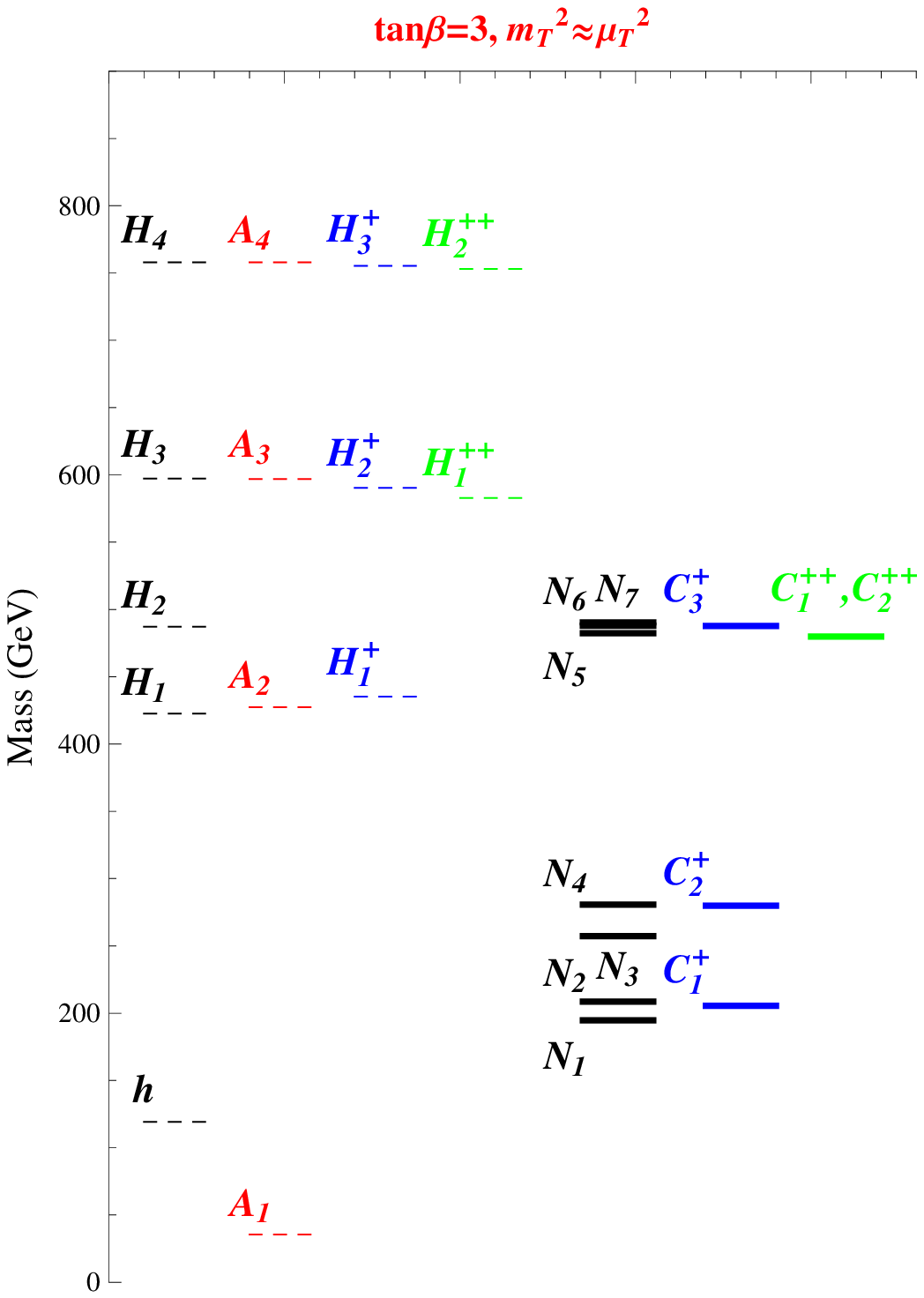}
\caption{Sample particle spectra for the three cases in the TNMSSM. The dashed and the solid lines denote 
scalars and fermions. $H$, $A$, $N$, and $C$ stand for CP-even/charged Higgses, CP-odd Higgses, neutralinos, 
and charginos, respectively (with superscripts indicating electric charges). The mass spectrum for the lightest CP-even
Higgs includes the one-loop radiative correction 
from stop (we assumed $m_{\tilde t_R} = m_{\tilde Q_3} =300$ GeV) , whereas the other masses are tree-level. The light state $A_1$ corresponds to the 
``$R$-axion'' which is further discussed in subsection~\ref{sec:raxion}}\label{fig:spectra}
\end{figure}

As an illustration we show in Fig.~\ref{fig:spectra} representative spectra of Higgses, charginos and neutralinos in all the three different cases that
we mentioned above.  In all these cases we take $200$ GeV and $220$ GeV for the $U(1)_Y$ and $SU(2)_L$ gaugino masses, 
respectively, and the other input parameters are tabulated in Table~\ref{tab:param}. Regarding the Higgs mass 
correction, we again include one-loop correction from the stop for the SM-like Higgs mass.
  \begin{table}[t]
\begin{center}
\begin{tabular}{|c|c|c|c|c|c|c|c|}
\hline
Items in GeV$^2$& $m_{H_u}^2$ & $m_{H_d}^2$ & $m_S^2$& $m_T^2$& $m_{\bar{T}}^2$& $\mu_T^2$\\
\hline
Non-SUSY limit & $-132^2$ &$177^2$&$-181^2$&$601^2$&$601^2$&$252^2$ \\
SUSY limit& $-245^2$ &$401^2$&$-404^2$&$312^2$&$312^2$&$570^2$\\
Intermediate case & $-212^2$ &$338^2$&$-340^2$&$472^2$&$472^2$&$480^2$ \\
\hline
SUSY parameters&\multicolumn{6}{|c|}{$\chi_u=-0.25,\;\chi_d=-0.55,\;\lambda=0.3,\;\kappa=0.3,\;\lambda_T=-0.6$ } \\
\hline
A-terms (GeV)&\multicolumn{6}{|c|}{$A_u=-0.1,\;A_d=0.1,\;A=0.9,\;A_\kappa=0.5,\;A_T=-0.1$ } \\
\hline
\end{tabular}
\end{center}
\caption{Parameter sets chosen for the three cases. } \label{tab:param}
\end{table}
Conforming to the usual convention, we denote scalars and fermions by dashed lines and solid lines,
respectively. States labelled $H$ correspond to CP-even or charged Higgses depending on their electric charge. In the same 
manner, the labels $A$, $N$, and $C$ correspond to CP-odd Higgses, neutralinos, and charginos respectively. 

Of course, we can choose the gauginos masses to be larger than what we assumed here, i.e., closer to the TeV scale,
since they do not (at least directly) enter the consideration of little hierarchy problem that we had so far.
Gauginos 
contribute to neutralino and singly-charged fermion spectra so that some of these
particles can be heavier than shown in Fig.~\ref{fig:spectra}.
However, in the
non-SUSY limit, $\mu_T$ is always at the weak scale and the triplet 
has an admixture in all three 
(i.e., neutral, singly-charged and doubly--charged) fermionic sectors,
so that (at least) one eigenvalue in all three sectors is always at the weak scale.
In the 
SUSY limit, $\mu_T$ is larger, but the $\mu$-term for doublets (which contribute to the neutral and singly-charged 
fermionic sectors) is still at the weak scale (based on the usual 
consideration of naturalness of the weak scale). 
Thus, 
(at least) one eigenvalue in the neutralino and singly-charged chargino sector is always at the weak scale in this case, but the
doubly-charged fermions are always heavier.

\section{Analysis of the Model}\label{sec:analysis}
\subsection{Electroweak Precision tests in the Models with triplets}
\label{sec:eptriplet}
One of the most stringent tests on new physics  models  comes from the EW
precision observables. 
The VEV of the electroweak triplet will modify relation between the masses of the $W$ and $Z$ bosons from the one 
in the 
SM and is thus constrained to be very small. 
On the other hand, once the MSSM doublet $H_u$ and singlet $S$ get VEVs, the $F$-term of the triplet inevitably 
leads to the tadpole term for the triplet 
\bea
|F_{\bt}|^2\sim \c_u\lambda_T H_u H_u S^\dagger T^\dagger.
\eea
So even in the case when triplet soft mass is not tachyonic this tadpole will result in the triplet VEV  of the order
\bea
\label{est}
\langle T\rangle\sim \frac{\c_u \mu_T v_u^2 }{(m_T^2+\mu_T^2)}
\eea
where $\mu_T$ is the triplet  effective $\mu$-term $\mu_T = \lambda_T v_S$.\footnote{There is also a tadpole for triplet
from the $F$-term of $H_{ u, d }$ which is proportional to $\mu$-term for doublets, but it is suppressed in the large $\tan \beta$ limit (which is our interest here).
See also reference \cite{DiChiara:2008rg} for a similar
discussion, including a detailed analysis of constraint from the $T$ parameter, in the model with {\em zero} hypercharge triplet.}
So we can see that the only way to 
accommodate the experimental data is to  make the total mass $\sqrt{\mu_T^2+m_T^2}$ of the scalar triplet large.\footnote{Note that demanding small triplet VEV does not necessarily mean tuning because this VEV is triggered by a dynamical tadpole (and
not by a tachyonic  mass). 
We will further justify this point using Eq.~(\ref{estimate}) below.} 
The limit where only $\mu_T$ is large corresponds to the supersymmetric integrating out of the triplet considered 
in~\cite{Dine:2007xi,Carena:2009gx} and discussed
in section II. This limit was shown to be viable and it can ameliorate the little hierarchy problem. 

As mentioned in section II, here we will be interested in a different region, when $\mu_T$ is of order soft masses or even smaller. This can 
give large corrections to the Higgs quartic for \emph{any $\tan \beta$} (as discussed
in section II), but we should of course check the constraint
from the $T$-parameter which we do in detail in this section.  A similar model with triplets but without singlet (TMSSM), where the $\mu$ terms for Higgses 
and triplets are just bare terms has very similar properties in the EW symmetry breaking sector. So, in 
order to 
understand better the qualitative features of the parameter space of the TNMSSM, we will first analyze the 
EW precision physics in the TMSSM. 

\subsubsection{EW precision in TMSSM}
\label{sec:ewtmssm}
\begin{figure}
\centering
\includegraphics[scale=1]{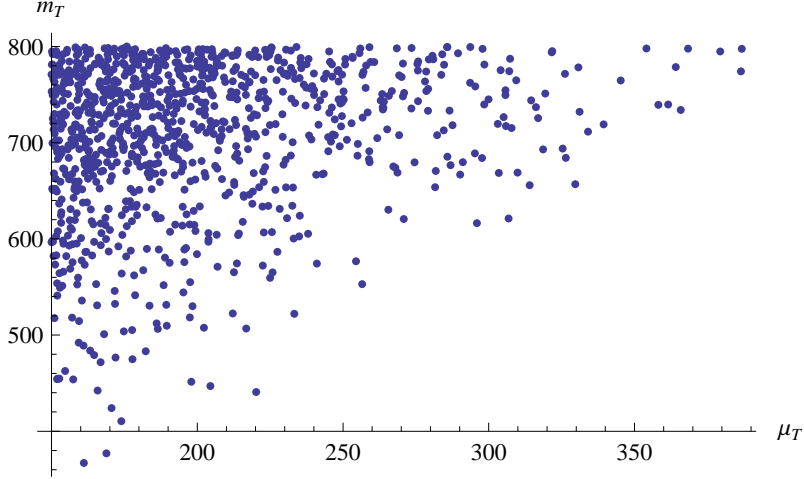}
\caption{
Sample viable points of the TMSSM 
in the plane of soft SUSY breaking ($m_T\approx m_{\bt}$) and supersymmetric mass terms
for triplets ($\mu_T$).
Every point on the plot has tree-level $T$ parameter consistent with 
data and \emph{tree-level} Higgs mass above 110 GeV. The other parameters (see text
for an explanation) are varied 
in the following ranges: $\c_{u,d} \in [-0.5,0.5]$, $\mu$ terms -
$\in [150,400]$
GeV, 
and $B\mu$ terms for doublets and triplets: $\in [-500^2,500^2]\hbox{GeV}^2$, $\tan\beta=10$.}
\label{tmssm_tree}
\end{figure}

We first review the bounds on the VEV of the triplet coming from $\rho$ parameter. For example,
the triplet with hypercharge $-1$ will have a VEV of the following form:
\bea
\langle T \rangle =\l(\baa{cc}
0&0\\
v_T&0\eaa\r).
\eea
Then the contribution to the Peskin-Takeuchi T parameter will be
\bea
\delta T & = &\frac{1}{\alpha}\frac{m_{W_1}^2-m_{W_3}^2}{m_W^2}=-\frac{1}{\alpha}\frac{2 v_{T}^2}{v^2} 
\eea
so that the constraint $T \gtrsim-0.1$ requires 
\bea
v_{T}^2 & \lesssim & (4 \hbox{ GeV} )^2
\eea
On the other hand we estimated in~(\ref{est}) triplet VEV in large $\tan \beta $ limit 
\bea
\label{estimate}
v_{T}\approx4  \times\frac{\l(\frac{\mu_T}{130}\r)}{\l(\frac{m_T}{600}\r)^2+ 
\l(\frac{\mu_T}{130}\r)^2}\l(\frac{\c_u}{0.4}\r) \l(\frac{v_u}{174}\r)^2
\eea
So we can see that with the soft mass of the order $\sim 600$ GeV\footnote{Further discussion of such a size of
soft mass term for triplet relative to those for doublets is in
section~\ref{sec:UV}.} 
and smaller $\mu_T$ (i.e., with only a mild hierarchy between the two triplet mass terms), 
we can accommodate the bound from the $T$ parameter.
In this case, the 
estimate for Higgs mass is given by Eq.~(\ref{eq:higgsupperbound}), except that we drop the singlet
contribution (2nd term) here. Thus, we obtain a large enhancement of the SM-like Higgs mass.
Another possibility for obtaining small $v_T$ is $\mu_T$ being larger than weak scale (and smaller $m_T$).
Here, the 
enhancement in Higgs mass is smaller: see Eq.~(\ref{SUSYtriplet}).
We have checked these estimates by numerical calculations 
shown in 
Figure~\ref{tmssm_tree} for $(\tan\beta=10)$,
where one can indeed see that points with small $\mu_T$ for triplets and large triplet soft masses
are preferred.
In fact, this plot is 
similar to that for the TNMSSM, i.e., adding the singlet, shown in Fig. ~\ref{fig:contour}
so that the
lesson here is the 
singlet is not so relevant for consideration of the little hierarchy problem and the $T$ parameter.

\subsubsection{One-loop analysis}
So far we have been focusing only on the tree level effects in EW precision tests arising from the VEV of the scalar triplet. Let us see 
now what will happen at one loop level. We know that the  new  fermionic and bosonic states will also contribute 
at the loop level to the 
$S$ and $T$ parameters. This contribution arises from the diagrams, where components of the triplet and doublet fields mix due to the Higgs VEV.
However in the non-SUSY limit that is our focus, the one loop contributions with doublet triplet scalar mixing are suppressed by the large 
soft mass of the triplets, but the same is not true for the fermions as follows. In order to 
maximize the increase in Higgs mass,
we need to be in the region of the parameter space with small $\mu_T$ so that generically 
contributions of the fermion loops with triplet fermion-higgsino mixing are important. 
The mass matrix for the neutralino fields (treated as $2$-compoment/Weyl spinors) will be given by:

{\tiny
\bea
&\hat{H}^TM_{mf}\hat{H}=\nonumber\\
&\l(\tilde H_u^0,\tilde H_d^0,\tilde T^0,\tilde\bt^0,\lambda_1,\lambda_2^3\r)\l(\baa{cccccc}
-2 \c_uv_{\bt}&-\mu & 0&-2\c_u v_u\tilde&\frac{g' v_u}{\sq}&-\frac{g v_u}{\sq}\\
-\mu &-2\c_d v_T& -2\c_d v_d&0&-\frac{g' v_d}{\sq}&\frac{g v_d}{\sq}\\
0&-2\c_d v_d&0&-\mu_T &\sq g' v_T&-\sq g v_T\\
-2\chi_u v_u&0&-\mu_T& 0& -\sq g' v_{\bt}&\sq g v_{\bt}\\
\frac{g' v_u}{\sq}&-\frac{g' v_d}{\sq}&\sq g' v_T &-\sq g' v_{\bt}& M_1& 0\\
-\frac{g v_u}{\sq}&\frac{g v_d}{\sq}&-\sq g v_T&\sq g v_{\bt}&0&M_2
\eaa\r)
\l(\baa{c}
\tilde H_u^0\\
\tilde H_d^0\\
\tilde T^0\\\tilde \bt^0\\
\lambda_1\\
\lambda_2^3 \eaa\r), 
\eea
} 
We need to know the couplings of these spinors in the mass eigenstate basis and in our analysis we will follow 
the discussion of Majorana spinors presented in~\cite{Bertolini:1990ek,Gates:1991uu}. The mass eigenstates will be related to the 
weak interaction eigenstates by orthogonal transformation $O$:
\bea
 \hat{H}^i=O^{i\alpha} N^\alpha,
 \eea
 where $N^\alpha$ are Majorana mass eigenstates, such that
 \bea
O^T.M_{mf}.O 
 \eea
is a diagonal matrix.
Then using properties of the Majorana fields one can show that the couplings of the mass eigenstates to the 
gauge bosons will be given by
\bea
&&A^\mu\bar{H^i_L}\gamma_\mu G_A^i \hat{H^i}_L=-\frac{1}{2}A^\mu\bar{N^\alpha}\gamma_\mu\gamma_5 N^\beta 
\l(O^T G O\r)_{\alpha\beta},\nonumber\\
&&G_{W_3}:~~~\frac{g}{2} {\rm Diag} (-1,1,-2,2,0,0),\nonumber\\
&&G_{B}:~~~~\frac{g'}{2} {\rm Diag} (1,-1,2,-2,0,0).
\eea 
where $G_{W,B}$ is gauge coupling matrix in the EW basis (for the left-handed fermion fields). 

\begin{figure}
\centering
\includegraphics[scale=0.7]{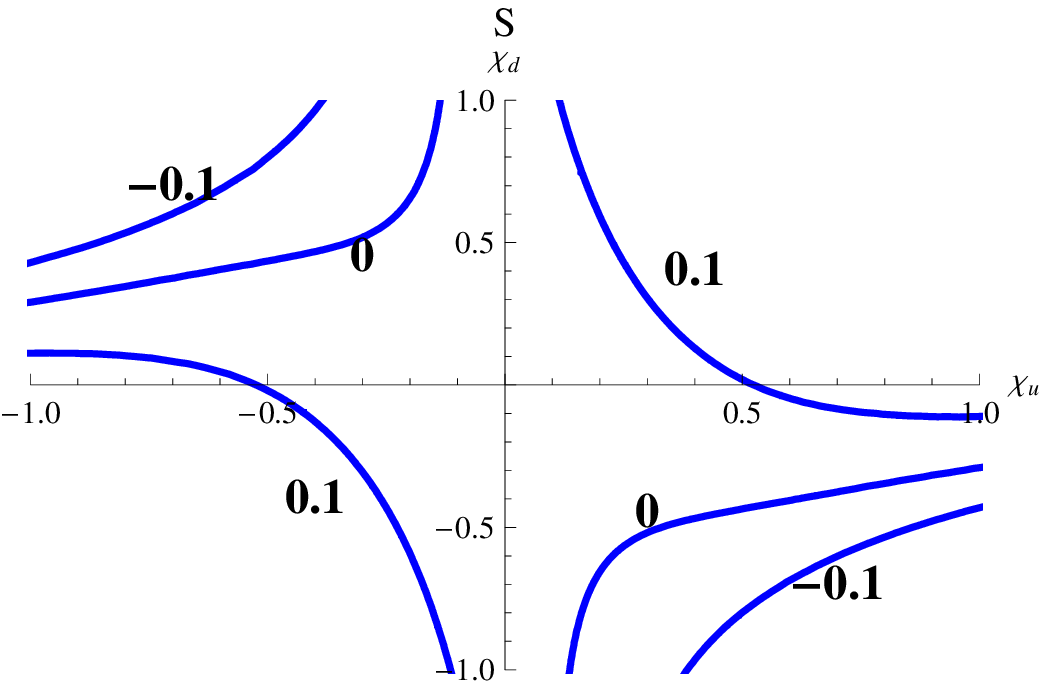}
\includegraphics[scale=0.7]{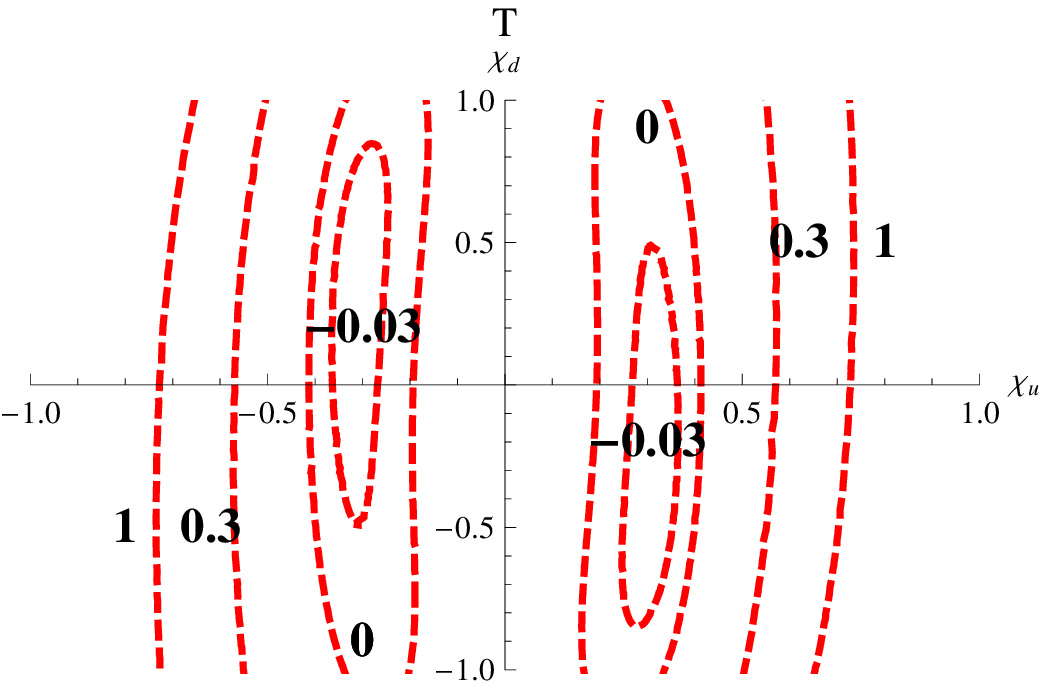}\caption{\label{M200}
Contour plot of $S$ (blue, solid)  and $T$ (Red,dashed) parameters in the $(\c_u,\c_d)$ plane
with $\mu_T=\mu=150$ GeV, $M_1=M_2=200$ GeV contributed by the fermions at one-loop
(see text
for an explanation of the parameters). 
}
\end{figure}

\begin{figure}
\centering
\includegraphics[scale=0.7]{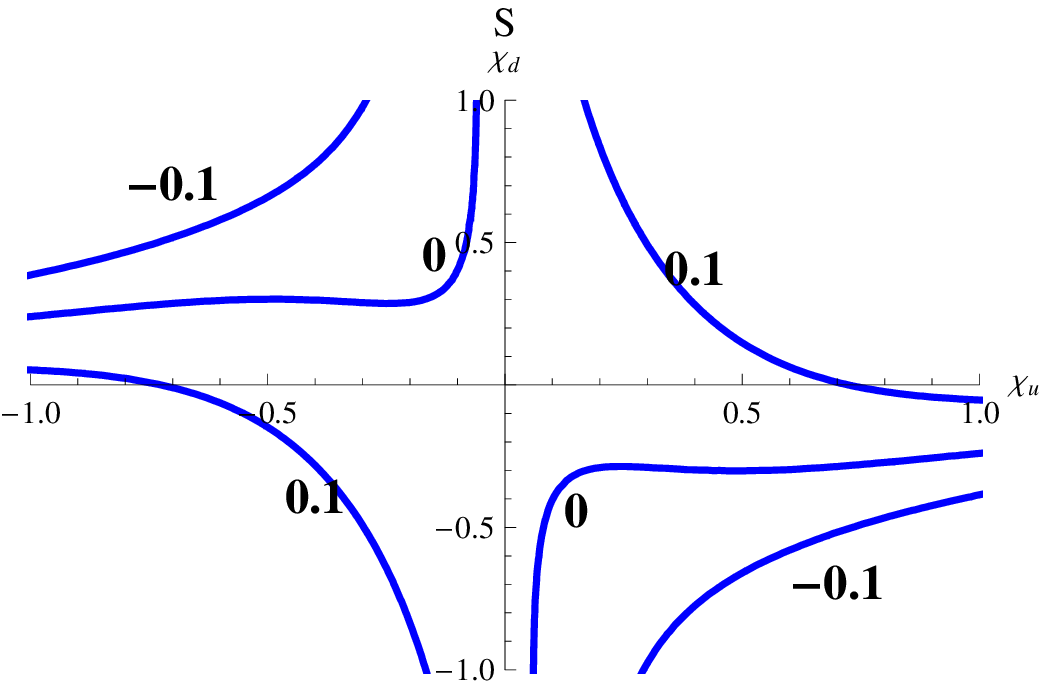}
\includegraphics[scale=0.7]{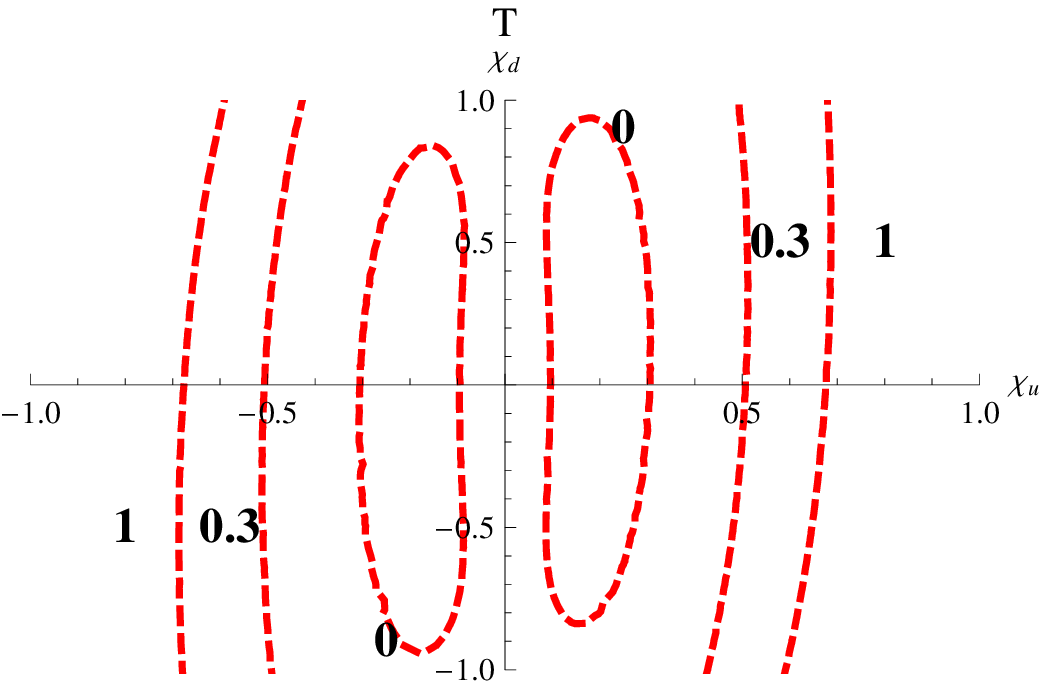}\caption{\label{M500}
Same as previous figure, except $M_1=M_2=500$ GeV}
\end{figure}

The charge-one fields have the following mass matrix:
\bea
(H_d^-, T^-, \lambda^-)_R\l(\baa{ccc}\mu & \sq \c_dv_d& g v_d\\
\sq\c_u v_u & \mu_T& \sq g v_{\bt}\\
g v_u&\sq g v_T&M_2
 \eaa\r)\l(\baa{c}
H_u^+\\
T^+\\
\lambda^+
\eaa\r)_L,
\eea 
These charge-one fermions have the following 
vector-like 
couplings to neutral gauge bosons:
\bea
B:~~~~\frac{g'}{2} {\rm Diag} (1,2,0)\nonumber\\
W_3:~~~\frac{g}{2} {\rm Diag} (1,0,2).
\eea
Similarly we can calculate the couplings of the fermions to the charged gauge bosons.
Now we calculate contribution of the higgsino-triplet fermion sector to the 
$S,T$ parameters. Results are presented on Fig.~\ref{M200},~\ref{M500}.  In these plots we have included only the 
contribution of the fermion sector . One can see that contribution to the $T$ parameter is almost always positive, 
compared to the tree level contribution which was always negative, so we will have some relaxation of the tree 
level bounds. Also it is interesting to note that a significant part of the parameter space predicts negative $S$ 
parameter, which relaxes EW precision bounds even more. 

Finally, we would like to mention the one-loop contributions of triplet might raise the Higgs mass even beyond the
tree-level effect, as calculated by reference \cite{DiChiara:2008rg} for the model with {\em zero} hypercharge triplet.
Since in our model it is rather easy for the tree-level Higgs mass to be beyond the LEP2 limit,
we defer such study of loop effects of triplet on Higgs mass for future work. We also notice that the contribution to the Higgs mass from
this effect is expected to be 
sub-dominant to that of the tree-level effects already discussed above.

\subsection{Light pseudoscalar}
\label{sec:raxion}
So far in our analysis we always assumed that soft SUSY breaking $A$-terms are small. This assumption was motivated 
by low scale gauge mediation which we considered as a possible UV completion of our model. As briefly discussed in 
section~\ref{sec:higgssector}, in the limit when all $A$ terms are zero
the potential is invariant under $U(1)_R$ symmetry
\bea
H_u,H_d,T,\bt,S\ra e^{i\phi_R}H_u,H_d,T,\bt,S.
\eea 
This symmetry is broken spontaneously by the VEVs of the $H_u,H_d, T ,\bt,S$
so that our spectrum contains a massless pseudoscalar, the  
pseudo-Nambu-Goldstone boson of spontaneous breaking of $U(1)_R$ symmetry of the model
(called the $R$-axion). 
This $U(1)_R$ is explicitly broken by the non-vanishing $A$-terms so that the mass of the $R$-axion will be suppressed 
by the value of the $A$-terms. In the limit where the soft masses of the triplets are large such that the VEVs of the 
triplet are small ($v_T,v_{\bt}\ra 0$), i.e., the triplet plays a negligible role in $U(1)_R$ breaking, 
the
$R$-axion is an admixture of the singlet and SM doublet, just like in the usual 
NMSSM:
\bea
\label{axion}
R_{axion}\approx\frac{v_s S_I+2vc_\beta s_\beta(c_\beta H_{uI}+s_\beta H_{dI})}{\sqrt{v_s^2+v^2\sin^22\beta}}.
\eea
In the limit when $A$-terms are small and $\tan\beta$
is large we will get the following expression for the axion mass: 
\bea 
\label{raxest}
m_{axion}^2
\approx -3A_S v_s-\frac{ 14Av_u v_d}{v_s}-\frac{2A_{\kappa} \kappa \lambda v_d(-8 v_s^2\lambda^2+
2v_u^2(g^2+{g'}^2+8\c_u^2))}{ v_s v_u(-8\lambda^4+4\kappa^2(g^2+{g'}^2+8\c_u^2))}.
\eea
As expected tree-level mass of the axion vanishes for the zero values of $A$-terms.

The couplings of R-axion to the SM fermions in the limit of large $v_S$,
large $\tan\beta$ and small values of triplet VEVs $ v_T,v_{\bt}$ 
are given by the same formula as in the NMSSM~\cite{Dobrescu:2000yn}:
\bea
\sim\frac{\sq}{v_s}\l( \frac{m_u}{\tan^2\beta}\bar{u}\gamma_5u+ m_d\bar{d}\gamma_5 d\r)iR_{axion}
\eea

In the case when $m_{axion}<m_\Upsilon$,
the decays of $\Upsilon\ra \gamma R_{axion}$ at $B$-factories can provide an
important test of the model, but it is possible to evade this bound by simply making the $R$-axion a bit heavier than $m_{\Upsilon}$.
Bounds from 
$Z \rightarrow R_{axion} \gamma$ are very weak  (see for 
example~\cite{Dobrescu:2000yn} and references therein)
due to the effective coupling involved in this decay arising from a loop of SM fermions,
combined with the suppressed nature of the $R$-axion couplings to the SM fermions (as above).  
The LEP searches for the $e^+e^-\ra Z^*\ra h A^0$ in MSSM~\cite{Schael:2006cr} can be reinterpreted as  searches for 
the light axion. However, these bounds are expected to be rather 
weak because the axion is always mostly singlet (see Eq.~\eqref{axion}).
In the large $\tan\beta$ limit,  
there is a further suppression due to the 
dependence on $\tan \beta$ in the admixtures of $H_{ u, d }$ in the
$R$-axion relative to those in the lightest CP-even Higgs.
Thus, in this limit, the 
$Z \left( h \dblarrow \d R_{axion}\right)$ coupling is estimated to be
\bea
\sim\sqrt{g^2+{g'}^2} \l(\frac{2v}{v_s}\frac{1}{\tan\beta^2}\r).
\eea
Numerical calculations show that this coupling is indeed is very small.
For a more detailed analysis  of the constraints on the light axion interactions in NMSSM, 
which applies to TNMSSM as well, we refer interested readers to~\cite{Ellwanger:2009dp,Dobrescu:2000yn}.

\subsection{Neutrino mass}
In the TNMSSM gauge invariance and renormalizability actually allow one more superpotential term other than the 
terms shown in Eqs.~\eqref{eq:model} and~\eqref{eq:yukawasuper}.
\bea
W= \xi L\cdot T  L \label{eq:badsup}
\eea
This is potentially dangerous because it could lead too large Majorana mass for neutrinos once the triplet with 
$Y=+1$ acquires its vacuum expectation value. Indeed, the Lagrangian contains the following mass term 
\bea
\mathcal{L}\ni \frac{1}{2}(2\xi v_T \psi_{\nu}\psi_{\nu}+h.c.),
\eea
i.e., the neutrino mass, which must be sufficiently small, is given by $2\xi v_T$. 

Even if we are required to have the triplet VEVs not more than $\sim 4$ GeV by the constraint from $\rho$ parameter 
(see section~\ref{sec:ewtmssm}), it would need an enormous tuning in order to suppress $v_T$ much below 
$\mathcal{O}(1\textnormal{ GeV})$, as required to approach the scale of neutrino masses:
see the issue of tadpole for triplets in section~\ref{sec:eptriplet}.  
We assumed that the coupling $\xi$ is $O(1)$ in the above argument. In order to avoid large neutrino masses, it is
of course technically natural to choose
this coupling 
$\xi$ to be very small. Indeed, we can 
forbid the superpotential given in Eq.~(\ref{eq:badsup}) by imposing a symmetry. 
One possibility is a $\mathbb{Z}_6$ under which the supermultiplets in the TNMSSM are charged as follows:
\bea
+1 \hbox{ for }L, \;\;\;+3 \hbox{ for }\bar{e},\;\;\;+2 \hbox{ for other fields}.
\eea
Even though this discrete symmetry is not anomaly free, one can think about it as an effective symmetry which holds up to very high energy scales. 

Under this symmetry, all terms in the TNMSSM that we had in earlier 
survive, except for the (unwanted) large Majorana mass term for neutrinos in Eq.~(\ref{eq:badsup}). 
Nevertheless, the (different) Majorana mass term, arising from the usual seesaw mechanism and thus
naturally highly suppressed: 
\beq
W\sim \frac{1}{M_s}(H_u L)^2
\eeq
is allowed.
Note that this symmetry forbids the usual renormalizable lepton-number violation operators
($W \ni LL \bar{e}$, $LQ \bar{d}$ and $L H_u$), allowing only the
operator $W \ni \bar u \bar d \bar d$. 
Thus, an intriguing possibility is that we do not impose $R$-parity, 
allowing the above baryon number violating term. Note that such baryon-number violating couplings
(i.e., appearing without the lepton-number violating terms) are relatively poorly constrained 
since they do not induce proton decay.
This would completely change the phenomenology of our model, but studying this possibility is 
beyond the scope of our paper.

\section{Evolution of parameters to high scales}
\label{sec:UV}
In previous section we discussed preferred values of TNMSSM parameters at the EW scale based on 
minimizing fine-tuning and EW precision tests. In this section,  we will 
RG evolve these parameters to higher energy scales in order to determine if 
there are any additional ``UV-considerations''. It is well known that the MSSM has two remarkable features 
when extrapolated beyond the EW 
scale: all its parameters stay perturbative up to very high scale (practically, the Planck
scale) and in fact, the three gauge couplings meet rather precisely at
$\sim 10^{ 16 }$ GeV. In this sense TNMSSM (by itself)
is less appealing since (as we show below) it lacks the
nice feature of gauge coupling unification. Nonetheless, we will show that
a simple modification of the TNMSSM has ``hints'' of unification. On top of that we show that all the couplings
which are important for the solution to the little hierarchy and $\mu$ problems 
that we presented here can be kept perturbative up to the ``GUT scale''.

Let us start with gauge couplings unification. When the triplets are added to the MSSM the one-loop $\beta$-function 
coefficients 
for the three gauge couplings are modified from the MSSM:
$(b_1, b_2, b_3) = (\frac{51}{5}, 5, -3)$ (we are using $SU(5)$ normalisation for the $g_1$ coupling).  With these coefficients, we find the $SU(2)_L\times U(1)_Y$ gauge couplings
meet around the 
``usual'' GUT-scale, i.e., $\sim 10^{ 16 }$ GeV. However the value of the gauge couplings at this scale ($\alpha \approx 1$) cannot quite be regarded 
as perturbative so that one-loop RGE equations might not suffice\footnote{Adding extra matter
charged under these gauge groups only makes the couplings more strong
at the GUT scale}. The $SU(3)_c$ gauge coupling does not unify with these two 
couplings, but it can attain a similar value (albeit large) at the GUT scale if 8 color triplets -- inert under $SU(2)_L\times U(1)_Y$\cite{Espinosa:1998re}
-- are added not far from the EW scale. Thus, even though perturbative/one-loop unification is lost in the TNMSSM, 
with a suitable modification, there is the possibility 
of a ``strong'' unification right below the Planck scale. 

Next, we consider the RG evolution of 
the new couplings (relative to the MSSM) which we introduced in order 
to address the little hierarchy problem and the $\mu$-problem of 
the MSSM, i.e., the couplings involving the singlet, triplet and the usual Higgs doublet fields.
A full set of relevant equations is given in 
appendix~\ref{app:RGE}. 
As seen there, all these couplings tend to grow in the UV due to contributions from these couplings themselves.
In addition, for the couplings involving Higgs doublets, the contribution of the (large) 3rd generation Yukawa couplings  
make matters worse here.
On the other hand, as is well-known, the (EW) gauge contributions have the opposite effect on the RG evolution of these
couplings.
The
point is that the Casimir  involved in these asymptotically free
effects is larger for the triplet couplings than for the others.

In the light of the above properties of the RG evolution, we 
expect $\lambda$ and $\kappa$ (the couplings of the NMSSM part of our model) to hit Landau poles before
the other couplings ($\chi_{ u, d }$ and $\lambda_T$) if they all have similar values at the weak scale.
However, note that we 
only need $\chi$'s to be relatively large
at the weak scale in order to solve the little hierarchy problem, i.e., 
we are not using the
$\lambda$ coupling to enhance the Higgs mass (unlike in the NMSSM). In fact,
we would like $\lambda$ to be relatively small since $\mu$-term for the doublets
($\sim \lambda v_S$) should be at the weak scale for naturalness. 
Thus, Landau poles should not be worry for our model.

\begin{figure}[t]
	\centering
	 \includegraphics[width=7.0truecm,height=7.0truecm,clip=true]{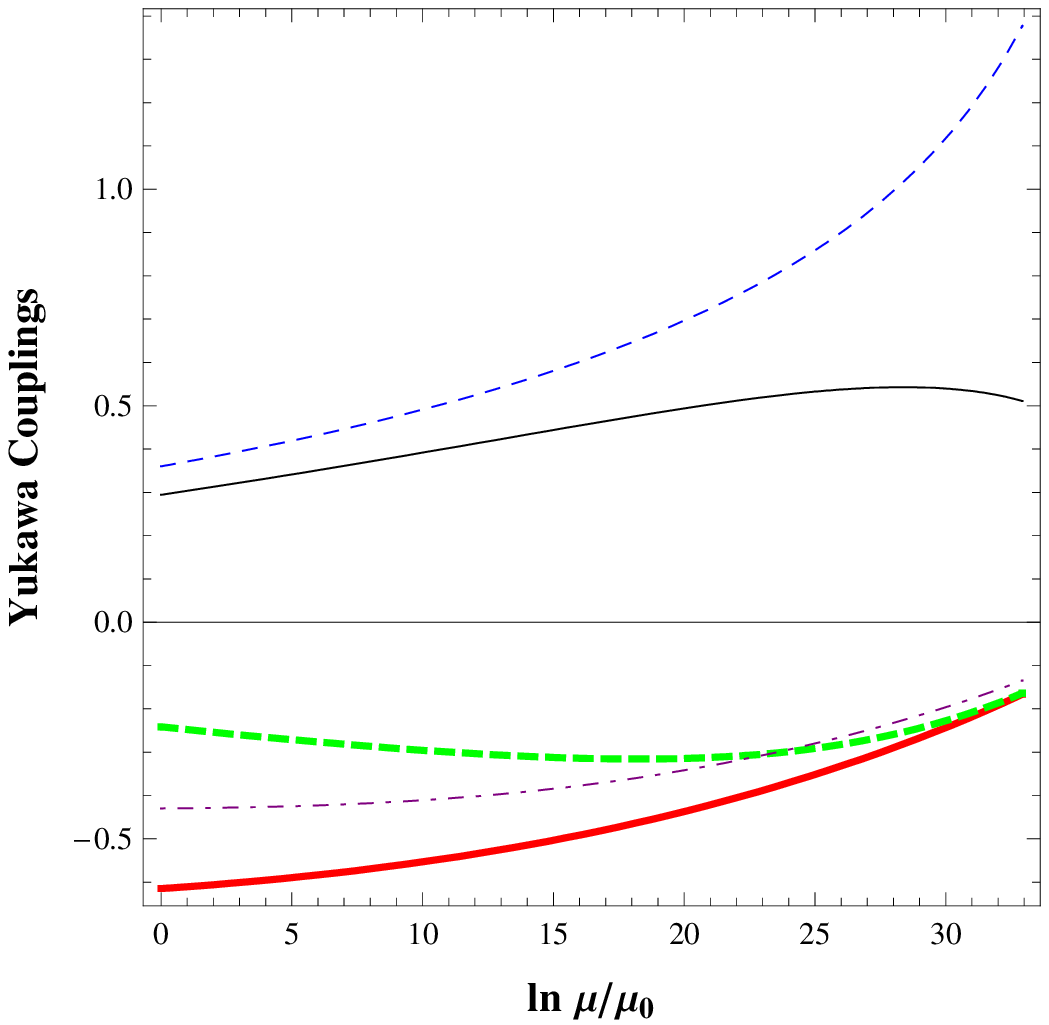}
	 \hspace{0.2cm}
         \includegraphics[width=7.0truecm,height=7.0truecm,clip=true]{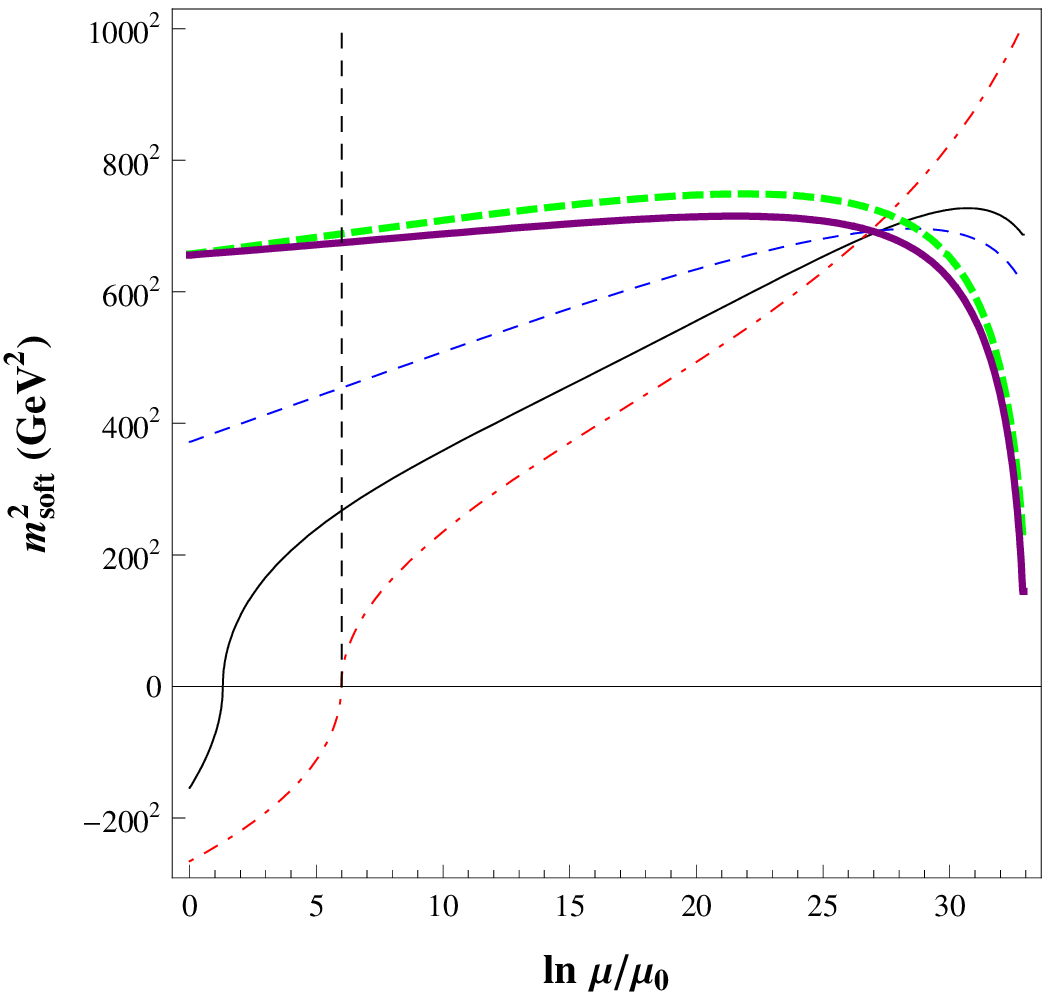}	
	\caption{On the left, evolution of Yukawa/superpotential couplings between the Higgses up to the GUT scale.  [Black/thin solid, Blue/dashed, 
	Red/thick solid, Green/thick dashed, Purple/dot-dashed] correspond (respectively)
	to [$\lambda,\kappa,\lambda_T,\chi_u,\chi_d$]. On the right, evolution of soft masses in the Higgs sector. [Black/thin solid, 
	Blue/dashed, Red/dot-dashed, Green/thick dashed, Purple/thick solid] correspond (respectively) to 
	[$m_{H_u}^2,m_{H_d}^2, m_{S}^2, m_{T}^2, m_{\bar{T}}^2$] and $\mu_0=100$ GeV. No new particles between the weak and the GUT scale are assumed in both cases.}
	\label{fig:rgyukawa}
\end{figure}

For illustration purposes we consider a sample point in parameter space, described in table~\ref{tab:values}. This point is fairly representative and one gets very similar results considering other values in parameter space consistent with naturalness and EW precision measurements.   
Running of the Yukawa couplings is depicted on the left panel of Fig~\ref{fig:rgyukawa} which confirms the above
expectations. 
In particular,  we can see that the values of $\chi$ required in order to enhance the
Higgs mass remain rather easily perturbative up to GUT scale.

\begin{table}[ht] 
\begin{center}
\begin{tabular}{|c|c|c|c |}
\hline
Gauge  & Yukawa & VEVs ($\tan\beta=5$)& $A$-terms \\
couplings & couplings & (GeV) & (GeV) \\
\hline 
&&&\\
$g_1=0.45$ & $\lambda=0.294$ & $v_u=170.6$ & $A=-7.48$ \\
$g_2=0.65$ & $\kappa=0.360$ & $v_d=34.1$ & $A_{\kappa}=1.46$ \\
$g_3=1.18$ & $\lambda_T=-0.615$ & $v_s=-519.4$ & $A_T=5.22$ \\
           & $\chi_u=-0.242$ & $v_T=2.85$ & $A_u=-2.25$ \\
           & $\chi_d=-0.430$ & $v_{\bar{T}}=-0.96$ & $A_d=4.62$ \\
           &&&$A_{h_t}=-335$ \\
           &&&$A_{h_b}=-40$ \\
           &&&$A_{h_e}=-45$ \\
&&& \\
\hline \hline
Gaugino & Soft mass & Soft mass & Soft mass \\
mass & (light scalars) & (heavy scalars) & (Higgses) \\
(GeV) & (GeV$^2$) & (GeV$^2$) & (GeV$^2$) \\
\hline 
&&&\\
$M_1=90$ & $m_{Q_1}^2=525^2$ & $m_{Q_3}^2=470^2$ & $m_{H_u}^2=-154^2$ \\
$M_2=100$ & $m_{u_1}^2=510^2$ & $m_{u_3}^2=390^2$ & $m_{H_d}^2=372^2$ \\
$M_3=570$ & $m_{d_1}^2=505^2$ & $m_{d_3}^2=500^2$ & $m_{S}^2=-266^2$ \\
 & $m_{L_1}^2=180^2$ & $m_{L_3}^2=180^2$ & $m_{T}^2=657^2$ \\
 & $m_{e_1}^2=115^2$ & $m_{e_3}^2=110^2$ & $m_{\bar{T}}^2=656^2$ \\
 &&&\\
\hline
\end{tabular}
\end{center}
\caption{Values of parameters at the weak scale for a sample point \label{tab:values}}
\end{table} 

Let us now briefly discuss the running of the soft masses in the Higgs sector. The running for the same reference point is illustrated on the right panel of Fig~\ref{fig:rgyukawa} (RGE's are again given in appendix~\ref{app:RGE}). 
Recall that we assumed that the soft mass$^2$ for $H_u$ and $S$ are negative at the weak scale
(as required for these scalars to acquire VEVs), but those for triplet are are positive (and of course similarly for squarks and sleptons).
A very attractive scenario would be that all these soft mass$^2$ RG evolve to positive (and roughly similar) 
values in the UV, i.e.,
radiative symmetry breaking (as happens for EW symmetry, i.e., $H_u$ in the MSSM).

From the right panel of Fig.~\ref{fig:rgyukawa}, we see that (not surprisingly) this indeed happens for $H_u$ just like in the MSSM.
What is more remarkable (compared to the usual NMSSM) is that singlet behaves similarly\footnote{Note that, in
the context of gauge mediation of SUSY breaking, 
the RG scale where singlet mass$^2$ vanishes can be taken to the messenger scale.}.
We would like to stress that this difference between the TNMSSM and the NMSSM
is due to the  interaction of the singlet with triplets giving an additional negative contribution to the running (from UV to IR)
of $m_{S}^2$. 
Thus, by (radiatively) generating a sufficiently large VEV for the $S$, the singlet-triplet coupling 
 significantly enlarges the viable parameter space.  Note that driving the singlet mass$^2$ negative this way does require the singlet-triplet coupling to be
larger  than $~\sim 0.1$. Combining this condition with $\mu_T$ being weak scale (or $\sim$ 100 GeV) implies that singlet VEV should
be less than $\sim$ 1 TeV.
On the other hand the soft mass$^2$ of the triplets undergo very moderate change in RG evolving to the UV, mainly because positive Yukawa contributions are largely compensated by negative terms proportional to the gauge couplings (which again come with 
a large
Casimir). This feature is crucial in
allowing for the possibility that the soft mass$^2$ for the triplet and doublet are (roughly) similar in the 
UV.\footnote{In the context of gauge mediation of SUSY breaking, the mild hierarchy between the soft masses in the UV, i.e.,
at the messenger scale, for triplet and doublets (see  
the right panel of Fig.~\ref{fig:rgyukawa}) can arise due to the larger Casimir for triplet vs. doublet.}

\section{Conclusions/Discussions}\label{sec:conclusions}

In this paper, we have presented an extension of the MSSM, based on adding 
$SU(2)_L$
triplet fields and a SM gauge singlet (as in the NMSSM) coupled to each other,
which solves the little hierarchy and $\mu$ problems of the MSSM. Our focus was on presenting a
complete model, performing a thorough analysis of electroweak precision tests and 
providing an origin for all mass scales in the model. We have started from a completely scaleless superpotential, such that the only scale in the problem is a soft mass scale, and showed that one can get a completely viable model 
which 
dynamically generates all the necessary scales, including effective $\mu$ and $B\mu$ terms
for the doublet and triplet fields.     

As the first step we reanalyzed the TMSSM (triplet-extended MSSM) as a good candidate for solving the little hierarchy problem. We explicitly showed that in this model the triplet inevitably gets a VEV (even if it is not tachyonic), 
which is of course severely constrained by the $\rho$-parameter. This problem has usually been circumvented in the literature by assuming a large $\mu$-term for the triplet and thus analyzing a model where these fields can be safely integrated out supersymmetrically, and all the interesting effects can be incorporated in the MSSM Lagrangian
augmented by higher-dimensional operators. This approach 
has the important drawbacks that the correction to the Higgs quartic is small in large $\tan \beta $ limit
and decouples with the $\mu$-term for triplet.

In this paper we took another approach, showing that the part of parameter space with relatively heavy triplet scalar and light triplet fermions (i.e., big soft masses, but small $\mu$-term) can be even more attractive.
It solves the little hierarchy problem for any $\tan \beta$ and even for very large soft mass for triplet. At the same time $S$ and $T$-parameters are well under control since the largest (potentially) dangerous contribution (to the $T$ parameter)
comes from the triplet scalar VEV
at the tree level, which however is suppressed
by the large soft mass. We also note that  adding a zero-hypercharge triplet, whose VEV gives an opposite contribution to $T$ parameter, can also potentially ameliorate a tension with EW precision tests~\cite{Georgi:1985nv,Chivukula:1986sp,Chanowitz:1985ug,Gunion:1989ci}. 

We then analysed a full singlet plus triplet extension of the MSSM (TNMSSM). We showed that coupling the singlet to the triplet 
has two major advantages. First, it naturally generates a weak-scale $\mu$-term for the triplet, just like the singlet-doublet coupling
of the NMSSM solves the doublet $\mu$-problem.
On the other hand, the triplet-singlet coupling helps to render the soft mass squared of the singlet negative along the RGE trajectories
thus enabling the singlet to acquire a VEV. In summary, we discover that the ``sum" of NMSSM and TMSSM is significantly more appealing than each of its components, taken separately.  

Finally let us comment on the issue of how the current LHC searches for SUSY might apply to this model . It is well known that these searches put very stringent bounds on squarks and gluinos below the TeV scale. However, these bounds heavily rely on several highly model-dependent assumptions. First, in order to put strong bounds on squark mass one needs the squarks of different
generations to be (roughly) degenerate in order to have big production cross sections. Superpartners are much harder to find if the third generation is somewhat special such that the stops and sbottoms are (much) lighter than rest of the squarks. It is also well known that one can ``hide" SUSY by squeezing the superpartner spectrum so that the
energy available to SM particles in superpartner decays is small. Usually this possibility is considered to be marginal. However in the TNMSSM one finds lots of new EW scale particles (including scalars and fermions) which are expected to be at the EW scale
thus making it easier to hide superpartners. To the best of our knowledge the bounds on these kinds of spectra are not well understood. Moreover, as we have already mentioned this entire scenario can be easily accompanied by $R$-parity 
(in particular baryon-number, but not simultaneously lepton-number) violation, which would significantly complicate the study. This would result in collider signatures without large missing transverse energy since the lightest SUSY particle would just decay into jets. Needless to say that such searches are much more difficult than standard SUSY searches and the current bounds on 
such a scenario are expected to be rather mild.    

It would be very interesting to understand better these bounds, as well phenomenology of our model in general. The latter can be of special interest (however, also possibly experimentally challenging) due to the enlarged neutralino, chargino and Higgs sectors.   
In particular, there is a light doubly-charged fermion (coming from the triplet) in the non-SUSY limit that
was our focus, unlike in the SUSY limit of this model or in the MSSM, NMSSM and extension of the MSSM with
{\em zero} hypercharge triplet.

\acknowledgments{We would like to thank  R.~Contino, J.~Evans, T.~Okui,  E.~Ponton
and R.~Sundrum for discussions. We are also grateful to Z.~Komargodski  for valuable comments on the manuscript. 
K.A. was supported in part by NSF grant No. PHY-0652363. A.K. was partially supported by NSF grant PHY-0801323.}

\appendix

%%%%%%%%%%%%%%%%%%%%%%%%%%%%%%%%%%%%%%%%%%%%%%%%%%%%%%%%%%%%%%%%%%%%%%%%%%

\section{Renormalization group equations} \label{app:RGE}
In this appendix we provide the RGE's in the $\overline{\textnormal{DR}}$ 
scheme for the parameters of the TNMSSM. The notations are $t=\ln(\mu/\mu_0)$ with $\mu$ the RG scale, 
and $g_2=g$, $g_1^2=\frac{5}{3}g^{'2}$ (with $e=g\sin\theta_W=g'\cos\theta_W$). 

Running of the new (relative to the MSSM) couplings is given by following equations:
\begin{eqnarray}
16\pi^2\frac{d\lambda}{dt}&=&\lambda\left[ 3h_t^2+3h_b^2+h_{\tau}^2+4\lambda^2+6\chi_d^2+
6\chi_u^2+3\lambda_T^2+2\kappa^2-3g_2^2-\frac{3}{5}g_1^2\right]\\
16\pi^2\frac{d\kappa}{dt}&=&\kappa\left[ 6\lambda^2+9\lambda_T^2+6\kappa^2\right] \\
16\pi^2\frac{d\lambda_T}{dt}&=&\lambda_T\left[ 2(\chi_d^2+\chi_u^2+\lambda^2+
\kappa^2)+5\lambda_T^2-8g_2^2-\frac{12}{5}g_1^2\right]\\
16\pi^2\frac{d\chi_u}{dt}&=&\chi_u\left[ 6h_t^2+2\lambda^2+14\chi_u^2+\lambda_T^2-7g_2^2-\frac{9}{5}g_1^2\right]\\
16\pi^2\frac{d\chi_d}{dt}&=&\chi_d\left[ 6h_b^2+2h_{\tau}^2+2\lambda^2+14\chi_d^2+\lambda_T^2-
7g_2^2-\frac{9}{5}g_1^2\right]
\end{eqnarray}

Running of the $A$-terms associated with the above new couplings is given by: 
\begin{eqnarray}
16\pi^2\frac{dA_\lambda}{dt}&=&A_\lambda\left[3h_t^2+3h_b^2+h_{\tau}^2+12\lambda^2+6\chi_d^2+6\chi_u^2+
3\lambda_T^2+2\kappa^2-3g_2^2-\frac{3}{5}g_1^2\right]\nonumber \\
&+&\lambda \left[6h_tA_{h_t}+6h_bA_{h_b}+2h_{\tau}A_{h_{\tau}}+12\chi_d A_d+12\chi_uA_u+
6\lambda_T A_T+4\kappa A_{\kappa}+6g_2^2M_2+\frac{6}{5}g_1^2M_1\right]\nonumber \\
&& \\
16\pi^2\frac{dA_{\kappa}}{dt}&=&3A_{\kappa}\left[2\lambda^2+3\lambda_T^2+6\kappa^2\right]+
3\kappa\left[4\lambda A+6\lambda_T A_T \right]\\
16\pi^2\frac{dA_T}{dt}&=&A_T \left[2\lambda^2+2\chi_d^2+2\chi_u^2+15\lambda_T^2+2\kappa^2-8g_2^2-
\frac{12}{5}g_1^2\right]\nonumber \\
&+&\lambda_T \left[4\lambda A +4\chi_d A_d+4\chi_uA_u+4\kappa A_{\kappa}+16g_2^2M_2+\frac{24}{5}g_1^2M_1\right]\\
16\pi^2\frac{dA_u}{dt}&=&A_u \left[6h_t^2+2\lambda^2+42\chi_u^2+\lambda_T^2-7g_2^2-\frac{9}{5}g_1^2\right]\nonumber \\
&+&\chi_u \left[12h_tA_{h_t}+4\lambda A +2\lambda_T A_T+14g_2^2M_2+\frac{18}{5}g_1^2M_1\right] \\
16\pi^2\frac{dA_d}{dt}&=&A_d\left[6h_b^2+2h_{\tau}^2+2\lambda^2+42\chi_d^2+
\lambda_T^2-7g_2^2-\frac{9}{5}g_1^2\right]\nonumber \\
&+&\lambda_t \left[12h_bA{h_b}+4h_{\tau}A_{h_{\tau}}+4\lambda A +2\lambda_T A_T+14g_2^2M_2+\frac{18}{5}g_1^2M_1\right]
\end{eqnarray}

%\subsection{Squark and slepton masses}
In order to describe the running of the soft masses, it is convenient to define following quantities:
\begin{eqnarray}
X_t &\equiv & h_t^2(m_{H_u}^2+m_{Q_3}^2+m_{\bar{u}_3}^2)+A_{h_t}^2 \\
X_b &\equiv & h_b^2(m_{H_d}^2+m_{Q_3}^2+m_{\bar{d}_3}^2)+A_{h_b}^2 \\
X_{\tau} &\equiv & h_{\tau}^2(m_{H_d}^2+m_{L_3}^2+m_{\bar{e}_3}^2)+A_{h_{\tau}}^2 \\
X &\equiv & \lambda^2(m_{H_u}^2+m_{H_d}^2+m_S^2)+A^2 \\
X_T &\equiv & \lambda_T^2(m_T^2+m_{\bar{T}}^2+m_S^2)+A_T^2 \\
X_u &\equiv & \chi_u^2(2m_{H_u}^2+m_{\bar{T}}^2)+A_u^2 \\
X_d &\equiv & \chi_d^2(2m_{H_d}^2+m_T^2)+A_d^2 \\
X_{\kappa} &\equiv & 3\kappa^2m_S^2+A_{\kappa}^2\\
S &\equiv & m_{H_u}^2-m_{H_d}^2+3m_T^2-3m_{\bar{T}}^2+tr[\textbf{m$_Q^2$-m$_L^2$-2m$_{\bar{u}}^2$+m$_{\bar{d}}^2$+m$_{\bar{e}}^2$}] \label{eq:definitionS}
\end{eqnarray}
where $\textbf{m}$ denote squark and slepton soft mass matrices in the generation space. 

The RG flows for the soft masses in the Higgs sector are then given by
\begin{eqnarray}
16\pi^2\frac{dm_{H_u}^2}{dt}&=&6X_t+2X+12X_u-6g_2^2M_2^2-\frac{6}{5}g_1^2M_1^2+\frac{3}{5}g_1^2S \\
16\pi^2\frac{dm_{H_d}^2}{dt}&=&6X_b
+2X_{\tau}+2X+12X_d-6g_2^2M_2^2-\frac{6}{5}g_1^2M_1^2-\frac{3}{5}g_1^2S \\
16\pi^2\frac{dm_T^2}{dt}&=&4X_d+2X_T-16g_2^2M_2^2-\frac{24}{5}g_1^2M_1^2+\frac{6}{5}g_1^2S \\
16\pi^2\frac{dm_{\bar{T}}^2}{dt}&=&4X_u+2X_T-16g_2^2M_2^2-\frac{24}{5}g_1^2M_1^2-\frac{6}{5}g_1^2S \\
16\pi^2\frac{dm_S^2}{dt}&=&4X+6X_T+4X_{\kappa}.\label{eq:rgms}
\end{eqnarray}
The running of the other Yukawa couplings (for example, that of top quark) and soft masses (for example,
that of stop) changes accordingly, i.e, taking into account the effect of the 
new couplings and soft masses.
We find that these changes are typically not important for our
purposes so that we do not provide here a complete list here. One can easily obtain 
these equations using the generic formula given in references \cite{Martin:1997ns,Martin:1993zk}.

%%%%%%%%%%
%%%%%%%%%%    References
%%%%%%%%%%

\bibliography{lit}
\bibliographystyle{apsper}

\end{document}